\documentclass[10pt,preprint]{aastex}

\usepackage{natbib}
\usepackage{lscape}
\usepackage{longtable}
\usepackage{amsfonts}

\usepackage{amsmath}
\usepackage[utf8]{inputenc}
\usepackage{multirow}
\usepackage{wasysym}
\usepackage{graphicx}
\usepackage[margin=1in]{geometry}
\usepackage{epstopdf}
\usepackage{psfrag}

 \makeatletter
 \g@addto@macro\normalsize{%
   \setlength\abovedisplayskip{0pt}
   \setlength\belowdisplayskip{0pt}
   \setlength\abovedisplayshortskip{0pt}
   \setlength\belowdisplayshortskip{0pt} }
 \makeatother

\begin{document}

 \title{OPTICAL - NEAR INFRARED PHOTOMETRIC CALIBRATION OF M-DWARF METALLICITY AND ITS APPLICATION}
 \author{\small{{\large{\bf {N. H}}}EJAZI, {\large{\bf { M. M. D}}}E {\large{\bf {R}}}OBERTIS}}
 \affil{Physics and Astronomy Department, York University, Toronto, ON M3J 1P3, Canada}
 \email{nedahej@yorku.ca, mmdr@yorku.ca}
 \author{\small{AND}}
 \author{\small{{\large{\bf {P. C. D}}}AWSON}}
 \affil{Physics Department, Trent University, Peterborough, Canada, K9J 7B8}
 \email{pdawson@trentu.ca}

 \begin{abstract}
Based on a carefully constructed sample of dwarf stars, a new optical-near infrared photometric calibration to estimate the
metallicity of late-type K and early-to-mid-type M dwarfs is
presented.  The
calibration sample has two parts; the first part includes 18 M dwarfs with  metallicities determined by high-resolution spectroscopy  and the second part contains 49 dwarfs with
metallicities obtained through moderate-resolution spectra. By
applying this calibration to a large sample of around 1.3 million M
dwarfs from the Sloan Digital Sky Survey and the Two-Micron All Sky
Survey, the metallicity distribution of this sample is determined and
compared with those of  previous studies. Using photometric parallaxes,
the Galactic heights of M dwarfs in the large sample are also
estimated. Our results show that stars farther from the Galactic plane, on average, have lower metallicity, which can be
attributed to the age-metallicity relation. A scarcity of metal-poor
dwarf stars in the metallicity distribution relative to the Simple
Closed Box Model indicates the existence of the ``M dwarf problem,"
similar to the previously known G and K dwarf problems. Several more
complicated Galactic chemical evolution models which have been
proposed to resolve the G and K dwarf problems are tested and it is
shown that these models could, to some extent, mitigate the M dwarf
problem as well.

\end{abstract}

 \keywords{galaxy: evolution - stars: late-type - stars: abundances - stars: fundamental parameters - techniques: photometric }

 \section{INTRODUCTION}

M dwarfs are the most numerous
stars in the Galaxy, contributing about 70$\%$ of all stars by number
(Reid \& Gizis 1997, hereafter RG97). Their main-sequence (MS)
lifetimes are much longer than the current age of the Universe and
they can therefore be used as excellent tracers of Galactic
structure and population as well as Galactic chemical, kinematical and
dynamical evolution. Since the advent of deep, advanced surveys such as
the Sloan Digital Sky Survey (SDSS, York et al.~2000) and the
Two-Micron All Sky Survey (2MASS, Skrutskie et al.~2006), M dwarfs
have been investigated in photometric and spectroscopic samples of
unprecedented size, revolutionizing this area of astronomy.

Clearly, a complete understanding of Galactic astronomy requires
accurate knowledge of fundamental properties such as mass, radius,
metallicity and temperature of these dwarfs. These,
 especially metallicity, however, have proven challenging to
calibrate. Although accurate values of metallicity for M dwarfs can
directly be obtained by analyzing high-resolution spectra (e.g., Woolf
\& Wallerstein~2005, hereafter WW05), the development of alternative
methods is needed. Since M dwarfs are among the intrinsically faintest stars , only a limited number of these stars are close enough for high-resolution spectroscopy (Woolf and Wallerstein 2006). For
this reason, efforts have been made to estimate the metallicity of M
dwarfs based on spectral band indices through
low-to-moderate-resolution spectra (L\'{e}pine et al.~2007, hereafter L07; Rojas-Ayala et al.~2010~$\&$~2012; Terrien et al.~2012; Mann et al.~2013a, hereafter M13a; Mann et al.~2013b, hereafter M13b; Newton et al.~2014, hereafter N14; Mann et al.~2014, hereafter M14). 

Despite the significant progress in deriving empirical metallicity calibrations using moderate-resolution spectra within the last few years, there is still a need for simpler approaches to determine the metallicity of large numbers of M dwarfs. There have been Several attempts to do this using
 photometric properties. By employing a calibration sample of M
dwarfs in M+FGK CPMSs, Bonfils et al.~(2005, hereafter B05) derived an
M-dwarf metallicity relation in terms of the {\it K}-band absolute
magnitude ($\mathrm{M}_{K}$) and $V - K$ color with a dispersion of
0.2 dex. Johnson and Apps~(2009, hereafter JA09) showed that while the
relation of B05 could reasonably reproduce the metallicity of
metal-poor M dwarfs, it underestimated the metallicities of their high
metallicity stars by an average of 0.32 dex. They established an
empirical model in which the distance of an M dwarf from the mean MS
 along $\mathrm{M}_{K}$ in the
$(V - K)$ - $\mathrm{M}_{K}$ plane was an indicator of its
metallicity. Schlaufman \& Laughlin (2010, hereafter SL10) suggested
that the  empirical photometric calibrations of B05 and JA09
systematically underestimated or overestimated metallicity at the
extremes of their ranges. They improved upon those calibrations and determined  the metallicity of an M dwarf using its distance from the
mean MS along the $V - K$ color (rather than $\mathrm{M}_{K}$) in the
$(V - K)$ - $\mathrm{M}_{K}$ plane. By applying a new sample of M dwarfs in
M+FGK CPMSs, Neves et al.~(2012, hereafter N12) showed that the
calibration of SL10 had a lower dispersion than those of B05 and AJ09,
and slightly modified this relation by readjusting its coefficients
using their own calibration sample.

In order to use the photometric calibrations described above, the
distances of M dwarfs are required; this limits their
applications only to samples of stars whose parallaxes are
available. A color-color diagram could provide a more
efficient technique for determining the metallicity of M dwarfs, even
those with unknown parallaxes. Several empirically determined metallicity
calibrations through color-color diagrams have been derived in the
last few years.  Using their large spectroscopic sample of M dwarfs,
West et al.~(2011, hereafter W11) obtained a two-dimensional fit which
related a metallicity parameter ($\zeta$, L07) to the $g - r$ and $r
- z$ colors, with typical uncertainties of 10 - 20\%. Bochanski et
al.~(2013) pointed out that the relation of W11 is limited only to
near-solar metallicity M dwarfs. They introduced a quantity,
${\delta_{(g-r)}}$ which measures the difference in $g - r$
between a subdwarf and its solar metallicity counterpart as a function
of $r - z$ color. Most recently, West et al.~(2014) used more than
20,000 M dwarfs with metallicities determined by the optimized
spectroscopic calibration of M13b (tested using a sample of wide,
low-mass binaries for which both components have an SDSS spectrum) and
derived relations between the metallicity and the SDSS {\it griz}
colors of M dwarfs. By employing an M-dwarf calibration sample in M+FGK CPMSs, Johnson et
al.~(2012, hereafter J12) developed a calibration which correlated the
metallicity of an M dwarf with its distance to the MS along the $J - K$
color in the $(V - K)$ - $(J - K)$ plane, with an RMSE = 0.15 dex. N14 assembled a sample of 447 M dwarfs
with metallicities calculated by their own spectroscopic calibration
and  found
that the $J - K$ color of an M dwarf is the best single-color
diagnostic of its metallicity. They then derived a metallicity relation
as a function of the distance of an M dwarf to the MS along $J - K$
color in the $(J - K)$ - $(H - K)$ color-color diagram, with a multiple
correlation coefficient ($R_{ap}^{2}$) of 0.92.

Our main goal in this study is to derive an optical-NIR photometric
metallicity calibration which can readily be applied to large numbers
of stars without the need for parallaxes, moderate-to-high resolution
spectra or  time-intensive calculations. This
will allow us to study statistically the metallicity distribution of the
local Galactic disk and test  models of Galactic chemical evolution
(GCE). The observations, sampling process and M-dwarf selection are
briefly described in section 2. Our metallicity calibration sample and
the best-fit relation for estimating M-dwarf metallicity are
presented in Section 3. The Galactic height ({\Large{\it{z}}} - height) distribution is
investigated in Section 4. Section 5 is devoted to the application of
the metallicity relation to our large sample and the comparison of the
resulting metallicity distribution with those from other studies. The statistical relation between
the metallicity and {\Large{\it{z}}} - height of M dwarfs is also discussed in
this Section. In Section 6, the Simple Closed Box Model (SCBM, Schmidt
1963) as well as a few more realistic GCE models are compared with our
M-dwarf metallicity distribution.

\section{OBSERVATIONS AND M DWARF SAMPLING}
  
\subsection{ Matched SDSS And 2MASS Data}
 The SDSS is one of the most extensive surveys in astronomy. During its
 operations, it obtained multi-color images which covered more than a
 quarter of the sky. Multi-band photometry was collected using a
 2.5-m wide-angle optical telescope at Apache Point Observatory, in
 New Mexico. The camera included  thirty CCD
 chips each with 2048$\times$2048 pixels (with a total of 120
 Megapixels). These chips were organized in five columns of six chips
 per column. Each column had a different optical filter bandpass,
 designated {\it u, g, r, i, z} (Fukugita et al.~1996), with average
 wavelengths of 355.1, 468.6, 616.5, 748.1 and 893.1 nm and with 95\%
 completeness of point sources in typical seeing to magnitudes of
 22.0, 22.2, 22.2, 21.3, and 20.5, respectively (Gunn et
 al.~1998). The SDSS imaging covers several thousand square degrees of
 sky, and over this region, photometric calibrations achieved an
 accuracy of $\approx$ 0.01 - 0.03 mag in {\it ugriz}
 (Ivezi\'{c} al.~ 2004,~ 2007; Adelman-McCarthy et al.~2006;
 Tucker et al.~2006; Padmanabhan et al.~2008).  
 
Our calibration uses both optical (the SDSS {\it g}) and NIR (the
2MASS {\it JHK}) broadband filter and we cross-matched the
SDSS and 2MASS data for collecting our sample.  The 2MASS was a
survey of the whole sky in the three IR broad-band filters {\it J, H}
and {\it K}, with average wavelengths of 1.25 $\mu$m, 1.65 $\mu$m and
2.17 $\mu$m, respectively. The measurements were made from 1997 to
2001 using two highly-automated 1.3-m telescopes, one at Mt. Hopkins,
Arizona for the Northern Hemisphere observations and one at Cerro
Tololo Inter-American Observatory, Chile for the southern hemisphere
data. Each telescope had a three-channel camera, each containing a 256
$\times$ 256 array of infrared detectors to scan the sky
simultaneously in the three filters.  Point sources brighter than
about 1 mJy, with S/N greater than 10 in each band were characterized
and compiled in a separate catalog, the 2MASS Point Source Catalog
(2MASS-PSC, Cutri et al.~2003). These bright sources ($\leq$ 13 mag)
generally have 1-sigma photometric uncertainty of $< 0.03$ mag
(Skrutskie et al.~2006).

We used the SQL Search tool in the SkyServer DR9 to collect an M-dwarf
sample suitable for our study. The stars were taken from the SDSS DR9
photometric catalog (identified by PhotoObj, Ahn et al.~2012) with the
object class specified by Type = 6 (for stars). Each object in the
SDSS DR9 catalog has a unique SDSS identifier, called ObjID, which is
also included in the 2MASS-PSC for objects in the regions that
overlapped with the SDSS. We therefore selected those stars with the
same ObjIDs in the two catalogs.

\subsection{Extinction Correction}

To minimize extinction, we chose stars with high Galactic latitude $ b
\geq 50\degr$. It has long been shown that, on average, extinction
decreases at higher latitudes, due to the low column densities
found along these lines of sight (Zagury~2006;Larson and
Whittet~2005).

We employed the dust maps of Schlegel et al.~(1998, hereafter S98) to
correct the stellar photometry for Galactic extinction.  Currently,
these maps provide the most comprehensive dust data of the Milky Way on
a large scale, including two two-dimensional full-sky maps (one for
the northern and one for the southern Galactic hemisphere) with total
line-of-sight dust column densities determined from far-infrared (100
and 240 $\mu$m) emission data. These maps have a resolution of 6.1
arcmin and are shown to predict reddening within 16\% (S98). 
We also used the relative extinctions given in Table 6 of S98 to convert
the extinction corrections in the $V$-band magnitude to SDSS-2MASS
filter bands. While the maps mentioned above provide an efficient tool
for estimating dust extinction, they refer to the total Galactic
extinction along the line of sight and so may overestimate the true
extinction to nearby stars (Covey et al.~2007 hereafter C07, Jones et
al.~2011 hereafter J11). Most of objects in the Galaxy lie behind 
only a portion of the Milky Way's total dust column and thus are likely
attenuated and reddened by only a fraction of the total dust column.

Using spectra of more than 56,000 M dwarfs from the SDSS, J11 created
a high-latitude, three-dimensional extinction map of the local
Galaxy. In their technique, spectra from  stars in the SDSS DR7
dwarf sample along low-extinction lines-of-sight were compared with
other SDSS M-dwarf spectra for deriving distances and accurate
lines-of-sight extinction. The three-dimensional map is most
appropriate for stars within around 500 pc from the Galactic plane ({\Large{\it{z}}} $\leq 500$ pc).  As mentioned in J11, it can safely be assumed that stars
with {\Large{\it{z}}} - heights greater than 500 pc are essentially behind the
entire dust column and the maps of S98 are more useful  for such stars.

\subsection{Clean Photometry}
Due to partial overlaps between neighboring SDSS images, an object
might be observed more than once. The best observation of an object is
called PRIMARY detection and the other observations, if any, are
assigned as SECONDARY detections. To avoid duplication, we chose only
PRIMARY observations by setting the variable ``mode" to 1. We also
used point-spread function (PSF) magnitudes for SDSS data since these
optimally measure the total fluxes for stars in this study.

By equating the flag CLEAN to 1, we selected those stars whose SDSS
magnitudes has passed appropriate standards of clean photometry (the flags: NOPROFILE, PEAKCENTER, NOTCHECKED, PSF-FLUX-INTERP, SATURATED, BAD-COUNTS-ERROR, DEBLEND-NOPEAK and INTERP-CENTER are not set). We
also set the 2MASS read flag to ``222", blend flag to ``111", and
contamination/confusion flag to ``000" to select only those stars
which have unsaturated, unblended and uncontaminated photometry in the
three IR bands ({\it J, H} , and {\it K}).

\subsection{Dwarf-Giant Separation}

We selected stars which fall within typical color ranges for late-type
K and M dwarfs, $r-i$ $\ga 0.47$ and $i-z$ $\ga 0.25$ (W11). There
possibly may be contamination by giants in any color-selected sample
of M dwarfs, however, which needs to be addressed in statistical
studies of the Galaxy. Bessell \& Brett~(1988, hereafter BB88) found
in the {\it JHK}, there is a clear bifurcation between M giants and M
dwarfs, beginning where TiO bands appear or the M-dwarf sequence
starts. By applying the NIR color ranges typical of M giants (BB88;
C07), we found that around 1\% of stars in our SDSS-2MASS sample are giants,
consistent with other studies such as  Covey et
al.~(2008, hereafter C08) with less than 2\% and W11 with around
0.5\% giant contamination.

It should be mentioned that this approach for separating giants from
dwarfs is not accurate for K-type stars (BB88) as they overlap in {\it
  JHK} color space.  A better way to remove giants employs reduced
proper motions of stars as described in M13b. By using a sample of
stars with available proper motions, giants could be separated from
dwarfs with higher accuracy.

\subsection{Photometric Parallax} 

The distance and {\Large{\it{z}}} - height of our stars were estimated by means
of a photometric parallax technique developed by Bochanski et
al.~(2010, hereafter B10). More specifically, the absolute magnitude
in the $r$ band, M$_r$, of low-mass stars can be obtained from a
$(J - K)-$ M$_r$ relation (the first equation of Table 4 in B10) which
is valid for the color range $0.50 < (r - z) < 4.53$. We therefore selected only those stars which fall in this color range.

After meeting all the criteria  above,
as well as the color cuts applicable to our metallicity calibration
outlined in Section 3, a sample of 1,330,179 M dwarfs was selected.

\section{METALLICITY CALIBRATION}
To calibrate metallicity using photometry, we collected a calibration
sample of M dwarfs with reliable metallicity and photometry. Since the
dust maps of S98 were used to remove Galactic extinction for our
large sample, the metallicity calibrations to be used for this  sample
must be based on calibrators with photometry corrected in the same
way. For this reason, we corrected the photometry of all stars in our
calibration sample using these maps\footnote{A dust map may make an offset in the extinction value of stars, specifically for each filter band. To minimize this effect on metallicity, the same offset must be included in the extinction of stars in both the calibration sample and large sample. For this reason, the same dust maps were applied for both samples.}.

There are nearby M dwarfs which are members of binary systems, having
 FGK-dwarf primaries with  metallicities determined through  high-resolution spectra.  We identified 14 such M dwarfs  with unsaturated SDSS $g$ and unsaturated
2MASS {\it JHK} magnitudes while  for most of these stars the photometry in the {\it r, i} or {\it
  z} bands are saturated. We used the SDSS${\_}$Phot Flag to
check the saturation status of stars; the  magnitude in a SDSS filter band is unsaturated
if the SATURATED flag for this band is not set. To be certain of unsaturated
{\it JHK} photometry, we chose only those stars whose 2MASS read flag
is ``222".

In order to add more stars with reliable metallicities to our
calibration sample, we refer to the work of Dahab \&
Strauss\footnote{Unpublished undergraduate Honor's thesis of
  W.E. Dahab under supervision of M.A. Strauss, Princeton University.}
(personal communication) who showed that by filtering the saturated
objects in SDSS that have certain flags set, the photometry of
remaining saturated objects are usable. The stellar locus on a color-color diagram can be identified by the locations at which photometrically clean stars appear. By comparing color-color diagrams of stars with unsaturated and saturated photometry, Dahab \& Strauss  demonstrated that the saturated PSF magnitudes in the {\it ugri} bands  for which  none of the
flags EDGE, INTERP-CENTER and PSF-FLUX-INTERP are set leave the color-color diagrams with almost no oulier. They also pointed out that  the {\it g} magnitudes of these saturated stars have a minimum of about 13.4 mag and the number of stars  fainter than this minimum increases exponentially. This indicates that saturated stars with {\it g} $<$ 13.4 mag are problematic and must be excluded from photometric studies.

\begin{figure}[h]
 \epsscale{0.8}
 \plotone{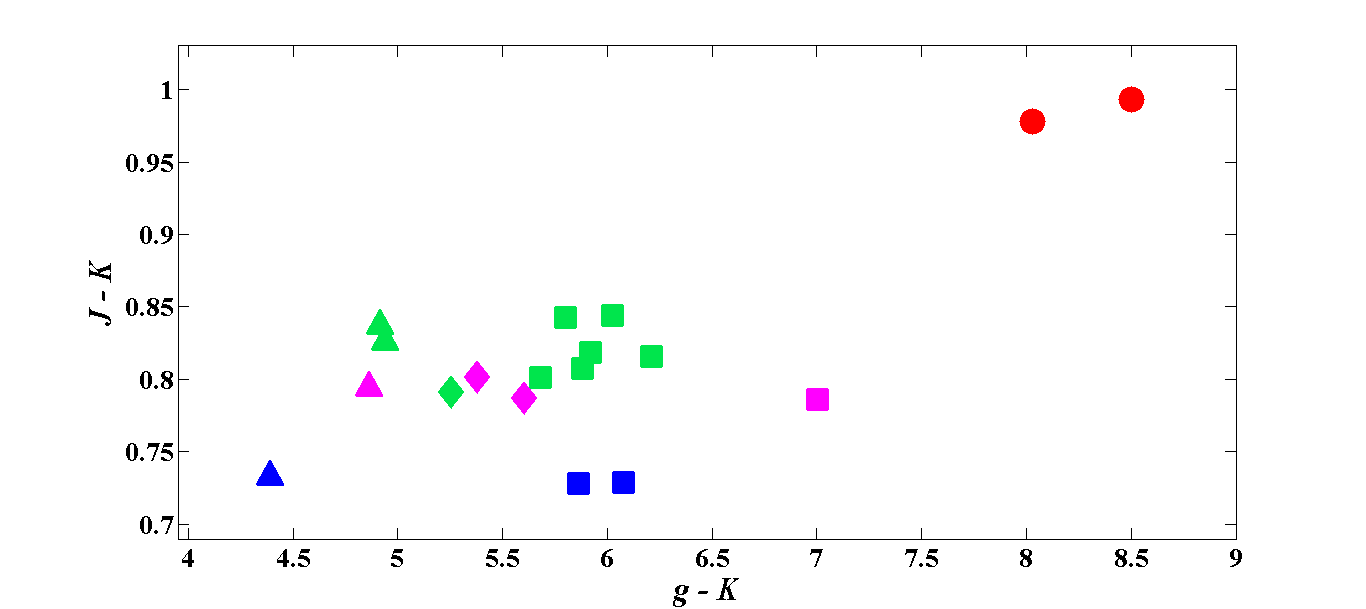}
 \caption{The  $(g - K)$ - $(J - K)$ color-color diagram for the 18 M dwarfs with metallicities determined by high-resolution spectroscopy.  The metallicity values are color-coded: stars with [Fe/H] $\geq$ +0.15 dex are plotted in red, with $-0.1$ $\leq$ [Fe/H] $<$ +0.15 dex in green, with $-0.4$ $\leq$ [Fe/H] $<$ $-0.1$ dex in purple and with [Fe/H] $<$ $-$0.4 dex in blue. The spectral types are symbol-coded: spectral types between around M0 and M2 are shown by closed triangles, around M3 by closed diamonds, between around M4 and M5 by closed squares, and around M6 by closed circles. }
 \end{figure}

\begin{figure*}[h]
 \epsscale{0.8}
 \plotone{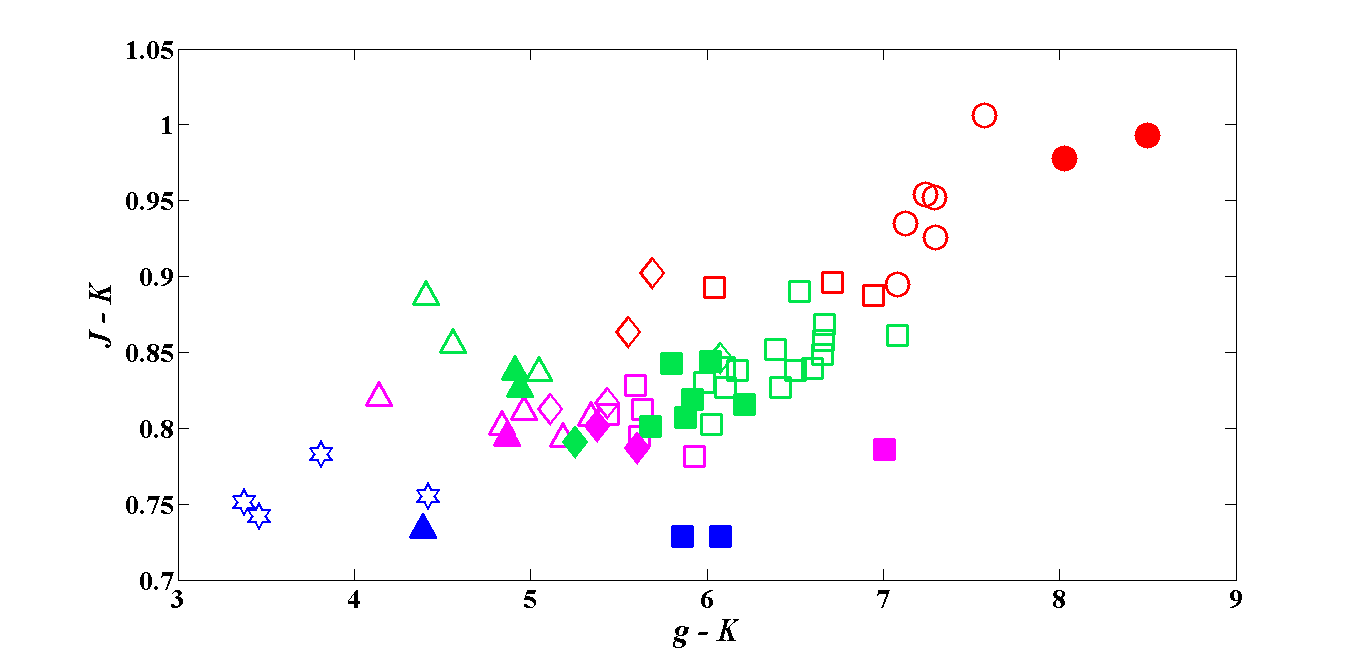}
 \caption{The $(g - K)$ - $(J - K)$ color-color diagram for the 67
   dwarfs in the calibration sample. The metallicity values are
   color-coded as described in Figure 2. The spectral types are
   symbol-coded: spectral types between around K6 and K7 are shown by hexagrams, between around M0 and M2 are shown by triangles, around M3 by  diamonds, between around M4 and M5 by  squares, and around M6 by  circles. The 18 stars from Figure 1 are depicted by closed symbols.}
 \end{figure*}

An object which is too close to the edge of an image is
flagged EDGE in the SDSS photometry. Among PRIMARY objects, only large
extended objects should be flagged EDGE.  Therefore, for point
sources, there is no need for concern about the PRIMARY objects having
the EDGE flag set. As a result, rather reliable photometry is possible
after filtering out the saturated objects in the SDSS whose
INTERP-CENTER and PSF-FLUX-INTERP flags are set. In
this way, we found three M dwarfs in M+FGK CPMSs, which have unsaturated
photometry in the {\it JHK} bands but whose {\it g} magnitudes are
saturated while none of the two flags are set. Moreover, the {\it g}
magnitudes of these three stars are  fainter (by around 4 mag) than
the lower limit  mentioned above, and we thus
call them ``weakly saturated" magnitudes. According to Dahab \&
Strauss, we can be certain that these magnitudes are reliable enough
for our study.

 The sample of M dwarfs in CPMPs now includes 17 (14+3) stars having metallicities determined through high-resolution spectra. Moreover, we found the measeured photometry and metallicity of LHS 3084  reliable and added this to the sample, increasing the number of stars to 18. Although LHS 3084 is not in a CPMP, its metallicity was determined by direct measurements on its high-resolution spectrum, not that of a primary. The astrometry, extinction-corrected photometry and  saturation status in the {\it g} band for the 18 M dwarfs are shown in Table 1. The spectral type of these  M dwarfs as well as the spectral type and metallicity of the primaries are given in Table 2. This Table also includes the references from which the metallicities and spectral types are taken.

The 18 M dwarfs described above form the foundation of our metallicity calibration. We investigated the relation between the metallicities and all  possible colors involving {\it g, J, H} and {\it K} magnitudes for these stars. It was found that the {\it J}$-${\it K}
color is the best metallicity indicator which is in agreement with the previous studies of
J12 and N14.  There are  deep potassium (K) and iron
hydride (FeH) absorption features as well as dozens of shallower metal
lines in the {\it J} band spectra of M dwarfs. Consequently (as J12 suggested), M dwarfs with higher
metallicity have preferentially suppressed {\it J} band fluxes as
compared to {\it K}-band spectra where there are only a few relatively
shallow Na and Ca lines as the prominent absorption features. This
causes the $J - K$ color of the metal-rich M dwarfs to be redder.

In addition, we found that among all possible color-color diagrams, the $(g - K)$ - $(J - K)$ diagram can reliably
separate metal-poor from metal-rich M dwarfs. Figure 1 shows such a
diagram for the  18 M dwarfs. The metallicity values are coded by
colors and the spectral types are coded by symbols provided in the
caption.  Overall, for a given spectral type range, metal-poor stars have
bluer $J - K$  than metal-rich ones, and clearly the $J - K$ color is a better
diagnostic for metallicity than $g - K$. It can also be seen that the
$g - K$ color is a better indicator for spectral type than $J - K$,
and for a given metallicity range, earlier-type M dwarfs have bluer $g - K$ than
stars of later spectral type. The location of a star in a color-color diagram theoretically  depends on its fundamental properties such as metallicity, spectral type (or temperature) and surface gravity\footnote{These are the three main properties  which play important roles in constructing model atmospheres and synthetic spectra of stars.}. However, as an approximation,  we assumed $\log(g)$ is constant for all M dwarfs ($\approx$ 5) and has  the same effect on the colors of these dwarf stars. Generally, different colors have different sensitivities to metallicity and spectral type.  In our case, while the $J - K$ could be a measure of metallicity,  the  $g - K$ is not sensitive enough to metallicity to separate metal-rich from metal-poor M dwarfs adequately. On the other hand, the $g - K$ color can distinguish different spectral types more effectively than the $J - K$ color.

Although the 18 M dwarfs provided us with a sample of robust calibrators, these stars by themselves are insufficient for covering the entire $(g - K)$ - $(J - K)$ plane, however,  as evidenced by gaps, particularly for $(g - K)$ values between 7 and 8. If we could add more dwarfs with trustworthy photometry and metallicity, there would be the same trends as those of the 18 M dwarfs: for a given spectral type range, more metal-poor stars should have smaller values of $J - K$, and for a given metallicity range, earlier-type stars should have smaller values of $g - K$\footnote{Similar to this can, to some extent,  be perceived from the work of N14 (see Figure 21 of their paper) on the $(J - K)$ - $(H - K)$ diagram.}. This defines a rule for selecting the rest of stars in our calibration sample. we collected 55 early-to-mid-type M dwarfs and late-type K dwarfs with spectroscopically determined metallicities
from moderate-resolution spectra. Their selection
initially was by eye, based on their locations on the
$(g - K)$ - $(J - K)$ diagram with respect to the original
18 M dwarfs. We chose those dwarfs which approximately followed the rule described above and this   enabled us to exclude some outliers from the sample in the first step of the selection. Assuming accurate photometry, large uncertainties in  metallicity and spectral type could show  a star in a metallicity-spectral type category different from its real one, causing it to significantly deviate from the rule.

We then obtained the best fit of a low-order polynomial between
metallicity and $g - K$ and $J - K$ colors for  the whole sample (18+55=73 stars) as
follows: 

\begin{equation}
[{Fe}/{H}] = C_1 + C_2 (g-K) + C_3 (J-K) + C_4 (g-K)^2 + C_5 (J-K)^2 + C_6 (g-K)(J-K) 
\end{equation}

\noindent
where $C_i$ are constant coefficients.  
This relation is of the same order (second order) as that of J12
who used an optical-NIR calibration with a filter set similar to
ours. The only difference is that we employed the $g$ magnitude
instead of the $V$ magnitude which was used in the calibration of J12. We
rejected 6 outliers for lying more than 2.5-sigma from the relation,
leaving a final calibration sample of 67 stars. The astrometry,
photometry, saturation status in the {\it g} band, spectral type, metallicity and related references for the
rest of our calibrators are listed in Tables 3 and 4. The $(g - K)$ - $(J - K)$ color-color diagram of the whole sample is presented in
Figure 2.

The coefficients of the final best fit for the 68 dwarfs in the
calibration sample are:

\hspace{3cm} $ {C_i} = \{-14.2959, 0.0519, 29.5926, -0.0529, -17.6762, 0.7032 \} $

\noindent                                          
with an RMSE ${=}$ 0.077 dex and ${R_{ap}^2}$ ${=}$ 0.90 (which are an improvement upon the values for the sample of  the original 18 M dwarfs: RMSE=0.090 dex and ${R_{ap}^2}$ ${=}$ 0.086), yielding
elliptical isometallicity contours. It should be remarked that this
metallicity calibration is applicable for stars of spectral types
between K6 and M6.5 with $-0.73$ $\leq$ [Fe/H] $\leq$ +0.3 dex, 3.37
$\leq g - K \leq$ 8.46 and 0.71 $\leq J - K \leq$ 1.01. Figure 3 shows
a comparison between the values of metallicity obtained by Equation
(1) with coefficients above and those taken from other studies for the
same 67 calibrators. We should mention that to have a more accurate metallicity calibration, a careful investigation of surface gravity is also required.

In the following sections, this metallicity calibration is applied to
our large sample and the resulting metallicity distribution is studied in more
detail.

\begin{figure}[h]
 \epsscale{0.8}
 \plotone{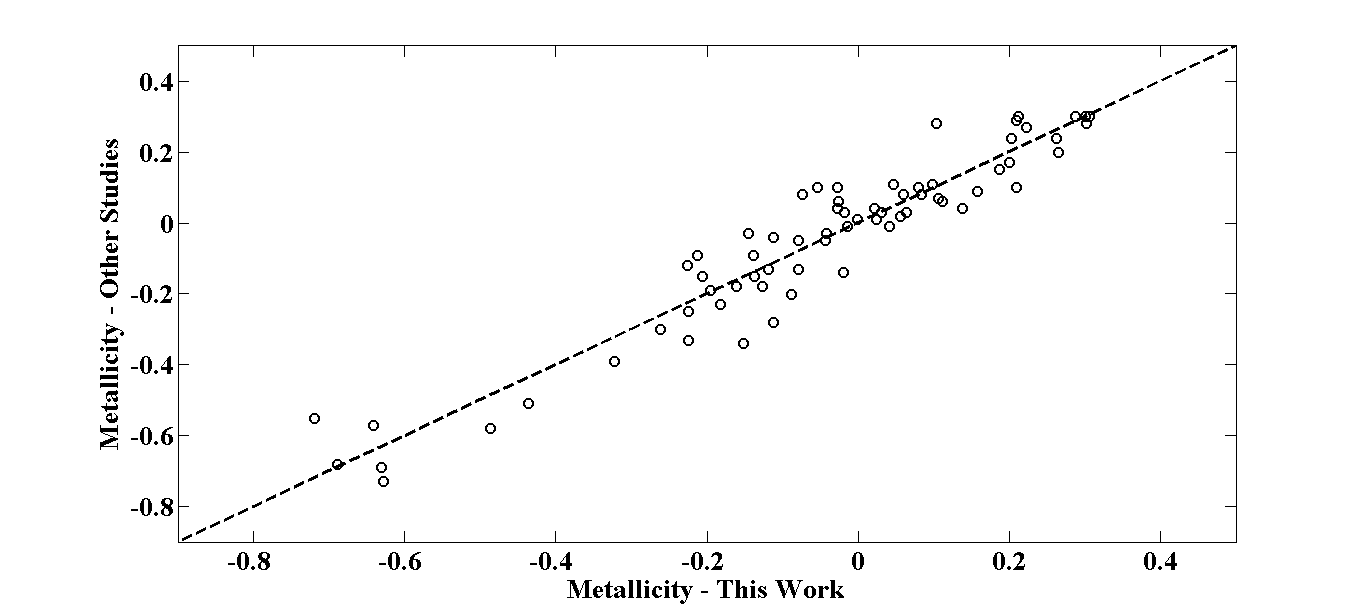}
 \caption{The metallicities derived by other studies versus metallicities calculated by Equation (1)                                                                                                                                                                                               }
\end{figure}

\section{{\Large{\it{z}}} - HEIGHT DISTRIBUTION}
To study the relation between metallicity and {\Large{\it{z}}} - height,  it is necessary to estimate the distance of stars in our sample. We applied Equation (1) to the sample of 1,330,179 M
dwarfs described in Section 2, accepting only those stars with the
metallicity range within which this equation is valid. This left a sample of 1,298,972 stars: only $\sim 1$\% of stars have metallicities outside this range; the rest
satisfy all requirements outlined in Section 2. We also limited our
study to stars with {\Large{\it{z}}} $\leq$ 2000 pc, leading to a final sample of
1,298,721 M dwarfs (hereafter, ``SampleMetal", or briefly ``SM"),
having a median extinction in the {\it r} band of $A{_r}=0.058$ mag.

We estimated the {\Large{\it{z}}} - height of stars in SM using the photometric
parallax described above (Subsection 2.5). It should be remarked that these
{\Large{\it{z}}} - heights are the vertical distances from the plane passing
through the Sun. Since all the selected M dwarfs lie above this plane
($ b \geq 50\degr$), we added the {\Large{\it{z}}} - height of the Sun,
{\Large{\it{z}}}$_{\sun}$ = 20 pc (Juri${\acute{\mathrm{c}}}$ et al.~2008), to those
  derived from the photometric parallax to obtain the distances
  from the central plane of the disk.
  
  The {\Large{\it{z}}} - height distribution of the sample (Figure 4) shows a peak at {\Large{\it{z}}} $\simeq$ 350 pc, indicating that around
94\% of the stars are within the Galactic thin disk, i.e., {\Large{\it{z}}} $\leq 1000$~pc
(Gilmore \& Reid 1983).

\begin{figure}[h]
 \epsscale{0.8}
 \plotone{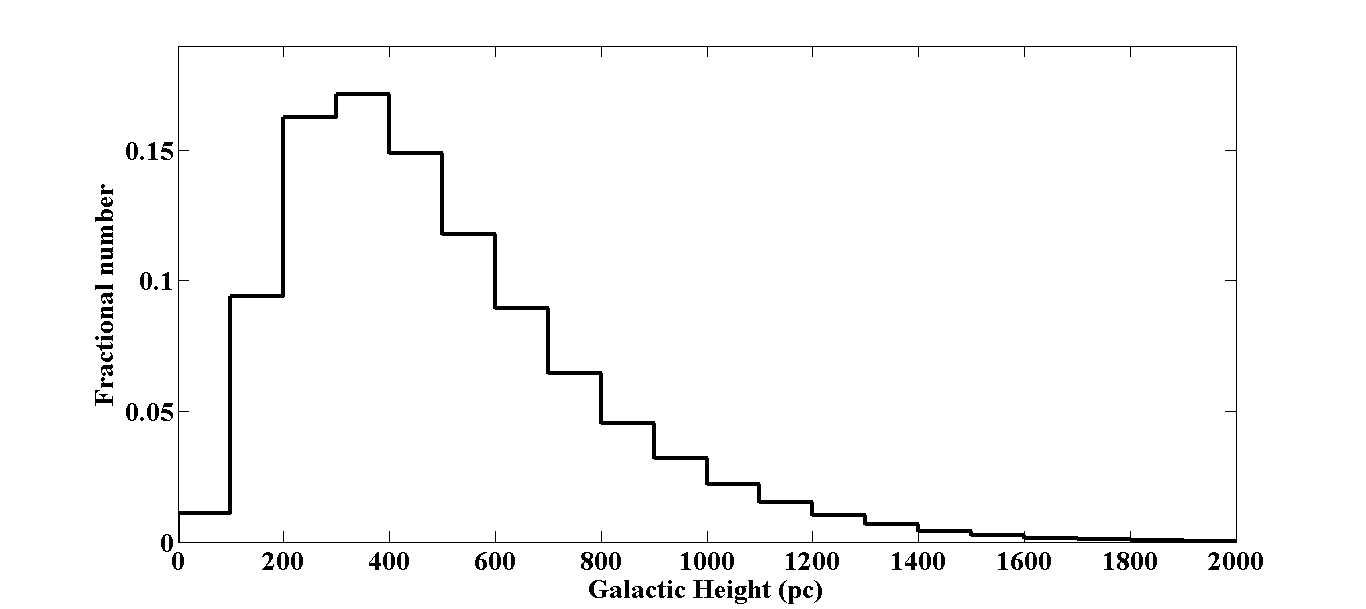}
 \caption{The {\Large{\it{z}}} - height distribution of stars in SM from the observed data}
 \end{figure}

\begin{figure}[h]
 \epsscale{0.8}
 \plotone{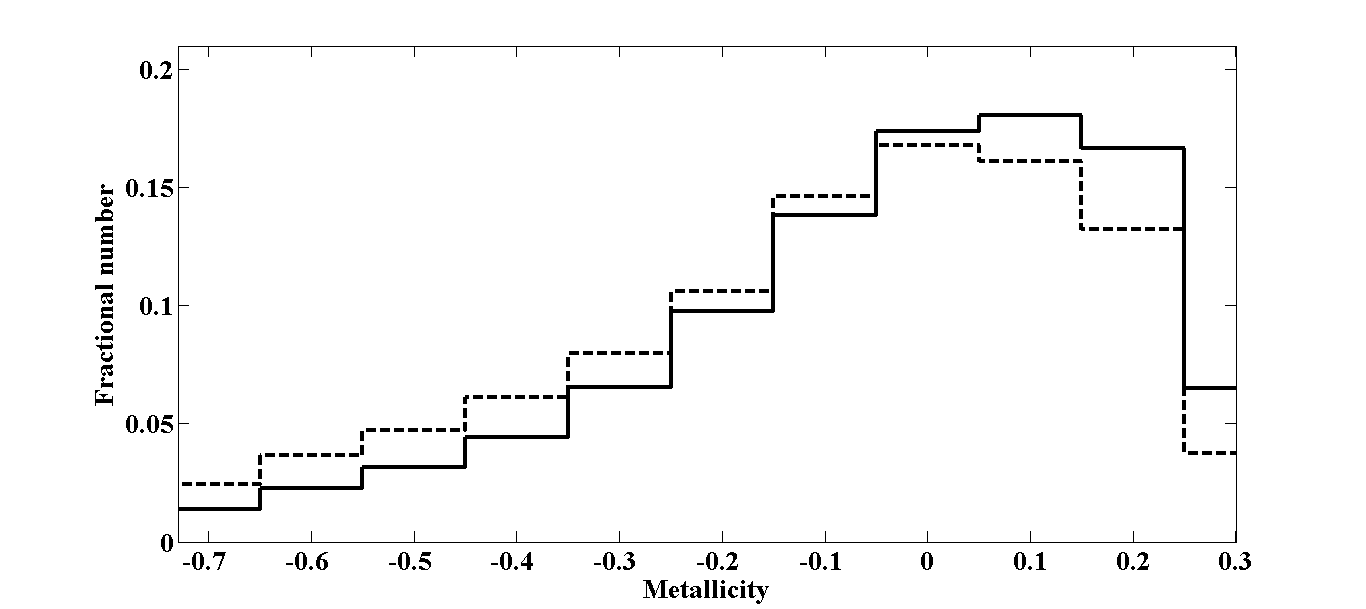}
 \caption{The metallicity distribution of stars in SM from the observed data (dashed histogram) and volume-corrected data (solid histogram) }
 \end{figure}

\section{METALLICITY DISTRIBUTION}
The metallicity distribution of stars from SM having $-0.7 \leq$ [Fe/H] $\leq +0.3$  is shown as a dashed histogram in Figure 5. As mentioned above, only those stars with $ b \geq 50\degr$ were selected. This means that stars outside an imaginary cone, perpendicular to the Galactic plane with an opening angle of $40\degr$ and apex  at Earth, were excluded. As a result, stars with higher vertical distances are distributed through larger volumes of space. In order to remove the effect of this bias on metallicity distribution, it is necessary to perform a
volume correction.  To this end, the stars in SM were divided into 20 bins of 100
pc in {\Large{\it{z}}} - height and the weighting factor
proportional to the inverse of the volume of each bin  was calculated. 
The metallicity distributions of these bins were obtained separately and  multiplied  by the corresponding weighting factors. All these weighted distributions were then added together as the total volume-corrected distribution, illustrated in Figure 5 (the
solid histogram). This  distribution shows a maximum at around +0.1 dex, and a 
mean metallicity [Fe/H] $\approx -0.04$ dex. Note that this mean
metallicity is not expected to represent the true mean value of the
local Galactic disk since those stars with metallicity outside the
metallicity range of our calibration were removed from the sample. In
order to obtain a more meaningful mean metallicity, a
calibration valid over a wider range of metallicity is necessary.

 \begin{figure}
 \epsscale{0.8}
 \plotone{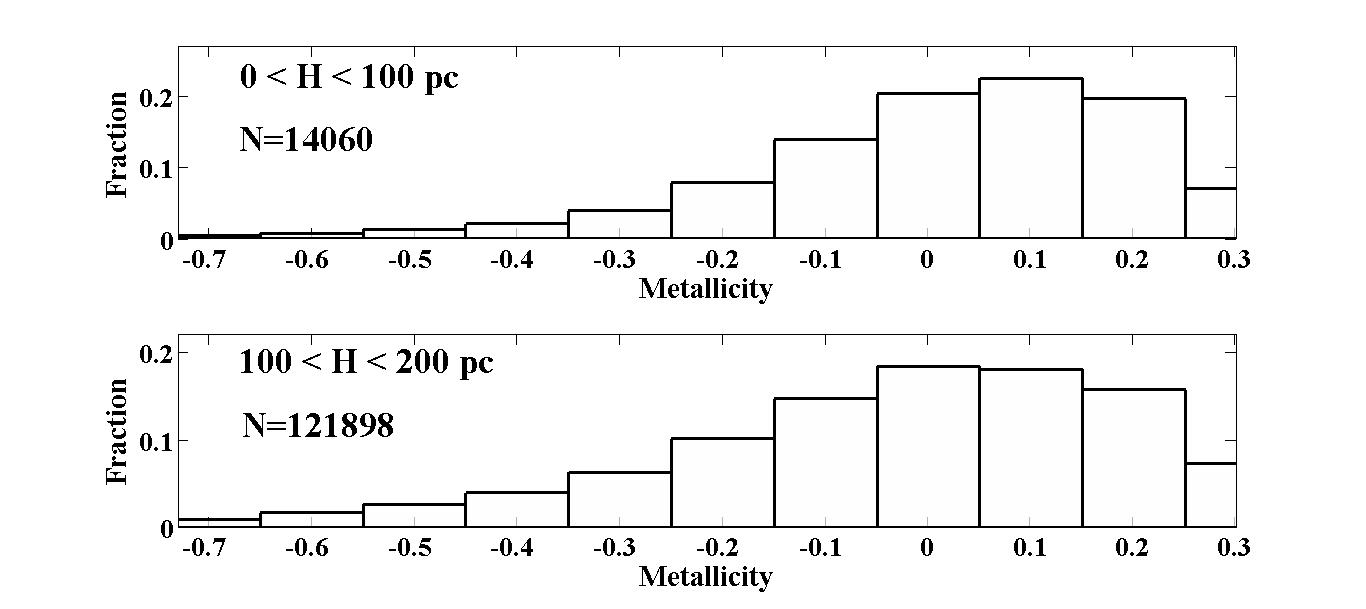}
 \plotone{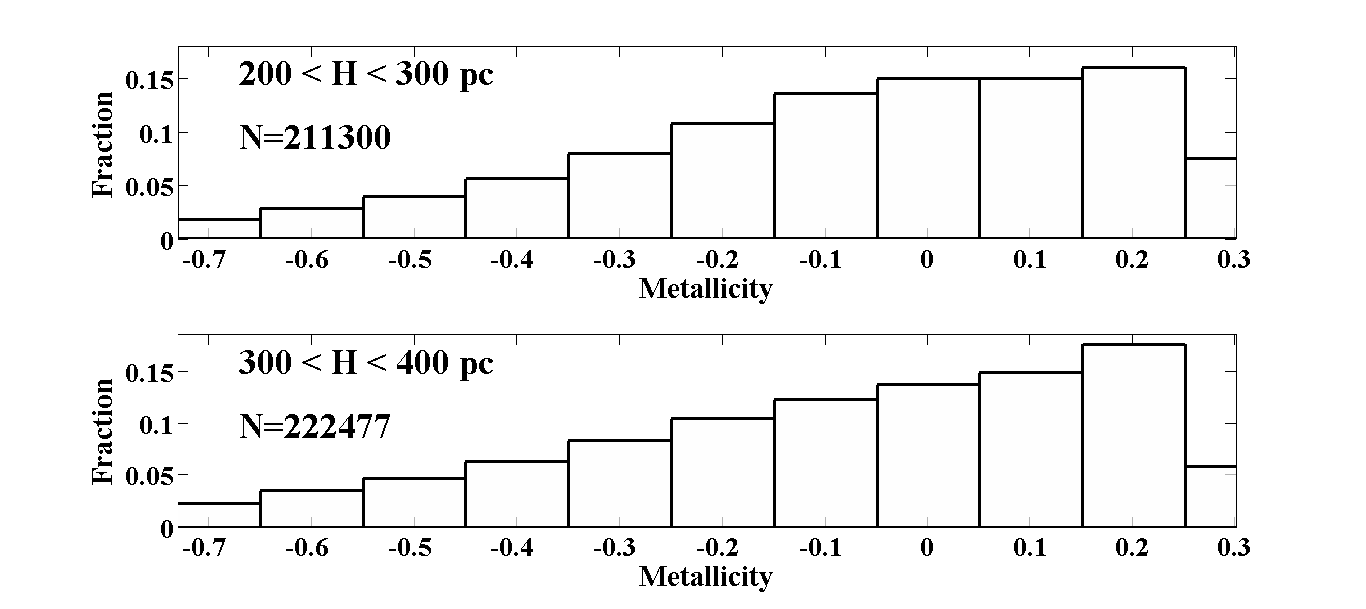}
 \plotone{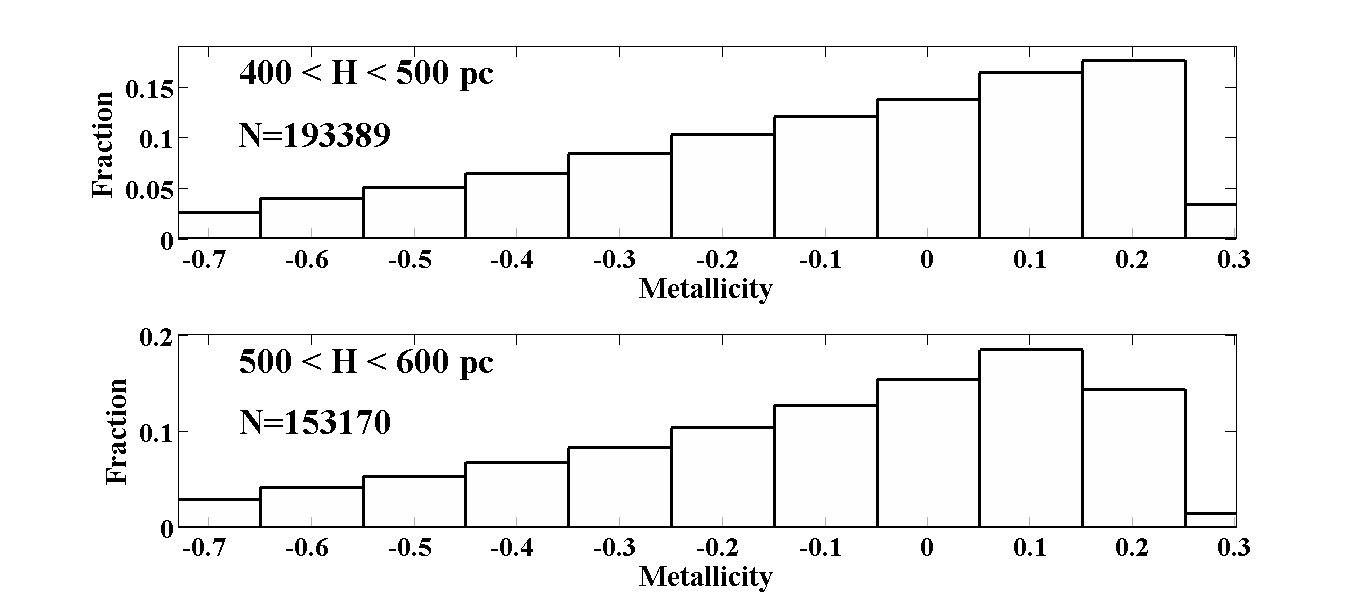}
 \caption{The metallicity distributions of M dwarfs in the large
   sample with different {\Large{\it{z}}} - height ranges between 0 and 600
   pc. The {\Large{\it{z}}} - height range and number of stars for each
   distribution are labeled on the corresponding plot. }
 \end{figure}

\begin{figure}
 \epsscale{0.8}
 \plotone{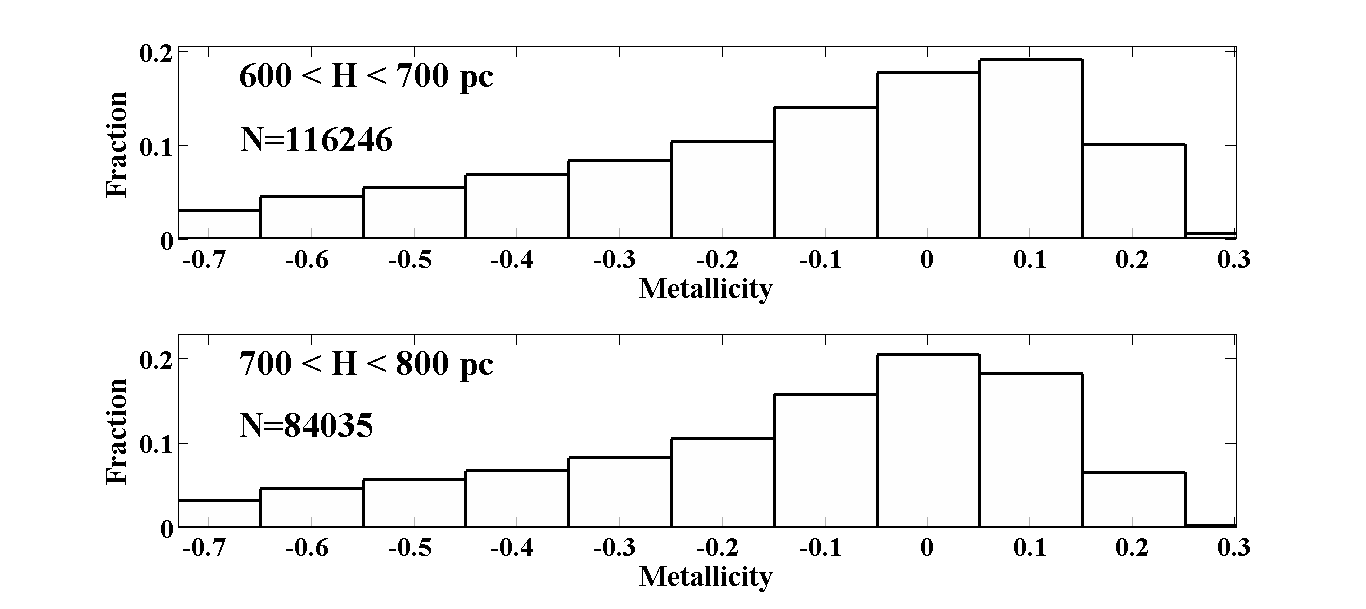}
 \plotone{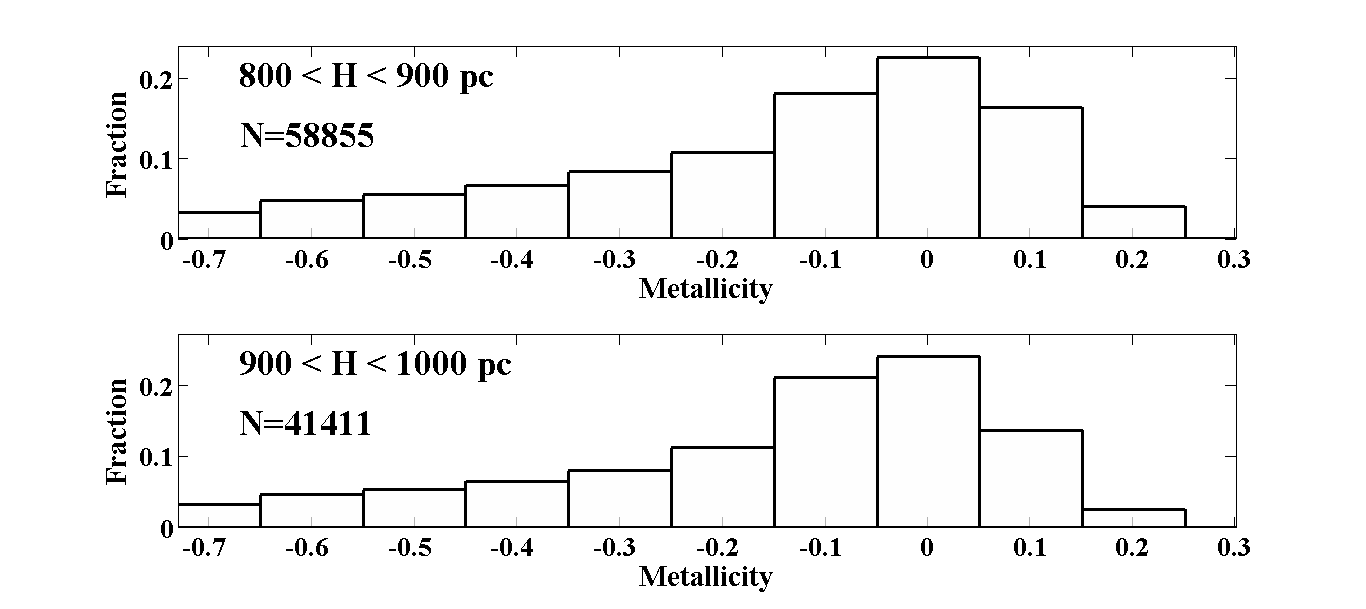}
 \plotone{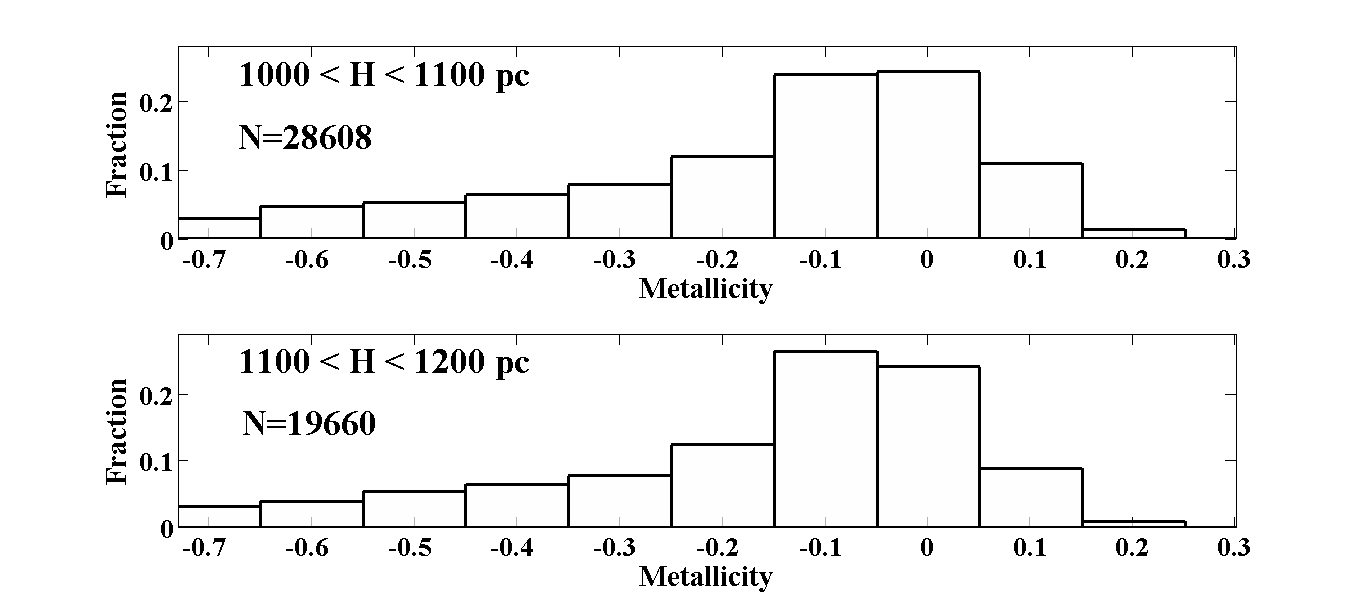}
  \caption{The metallicity distributions of M dwarfs in the large
   sample with different {\Large{\it{z}}} - height ranges between 600 and 1200
   pc. The {\Large{\it{z}}} - height range and number of stars for each
   distribution are labeled on the corresponding plot. }
\end{figure}

\subsection{Metallicity and {\Large{\it{z}}} - height Relation}
The relation between metallicity and distance {\Large{\it{z}}} above the Galactic plane  can be investigated by determining the metallicity distribution of stars as a
function of {\Large{\it{z}}} - height.  We divided a subsample of our stars with
{\Large{\it{z}}} $\leq$ 1200 pc into 12 bins of equal $\Delta${\Large{\it{z}}} = 100 pc.
The metallicity distribution associated with each bin is shown in
Figures 6 and 7. There is a clear shift towards lower
metallicities as {\Large{\it{z}}} - height increases.; the fraction of metal-rich stars slowly decrease and the distributions become more metal-poor  with increasing height.   The mean metallicity decreases (by around 0.19 dex) from 0 to 1200 pc which is consistent with the age-metallicity-height relation. Studies of
Galactic evolution have demonstrated that stars which formed at
earlier times in the Galaxy's history generally have lower metallicities in
general and are, on average, farther from the Galactic plane. Stellar systems are formed from
interstellar gas and, at the end of their lifetimes, they may return a
substantial fraction of their initial masses enriched with metals to the interstellar
medium. As a result, succeeding generations of stars become more metal-rich than
their predecessors, leading to an age-metallicity relation. The pioneering work on this was done by Twarog (1980a, 1980b) who derived the age-metallicity relation for the disk in the neighborhood of the sun using a large sample of southern F dwarfs. They found that the mean metallicity of the Galactic disk increased by about a factor of five between 12 and 5 billion years ago and has changed only slightly since then.   A recent evidence of this
kind was offered by Casagrande et al.~(2011, hereafter C11) who determined the
metallicity distributions of three samples of solar-type stars with
different ranges of age (Figure 16 of their paper). They showed
that young stars (ages $< 1$ Gyr) have a quite narrow distribution
around higher values of metallicity, whereas intermediate-age (between
1 and 5 Gyr) and old-age ($> 5$ Gyr) stars present broader
distributions, significantly extending to lower
metallicities.

In addition to the age-metallicity relation, there is also a relation
between age and velocity dispersion. Observations have demonstrated
that old stellar populations in the disk have larger velocity
dispersions. Based on the sample of Twarog above, Carlberg et al. (1985) examined the age-velocity dispersion relation for stars in the solar neighborhood. They found an increase in  velocity dispersion  to a maximum at about 6 Gyr, thereafter, the trend remained roughly constant\footnote{This behavior is consistent with a simple model for the local-disk formation and evolution in which growth occurs at a uniform rate with no initial disk, and the random velocities of stars are increased by a mechanism which depends on the total surface density of the disk, such as spiral waves.}. The term ``dynamical heating" is commonly applied to all 
processes that cause an increase in velocity dispersion with
age. Different mechanisms are responsible for vertical dynamical
heating (e.g., Nordstr\"{o}m 2008). Whatever the mechanism(s), stars
dissipate from the Galactic plane, increasing their vertical distance
from the plane in the course of time. One therefore expects that
older, more dynamically heated stars should be found, on average, at
larger {\Large{\it{z}}} - heights than the younger stars. As a consequence,
stars farther from the Galactic plane should be, on average, more
metal-poor than those closer to the plane.

Using a metal-sensitive ratio, (CaH2 + CaH3)/TiO5, from
L\'{e}pine et al.~2003 as a proxy for M-dwarf metallicity, West
et al.~(2008) demonstrated a decrease in metallicity as a function of
{\Large{\it{z}}} - height (up to 1000 pc) which implies that stars more distant
 from the Galactic plane have lower metallicities. In a
simulation by West et al.~(2006; 2008), a simple one-dimensional model
of the thin disk was applied to investigate the vertical motions and
positions of M dwarfs over the Galaxy's lifetime. The model also showed a
decrease in activity fractions (as traced by H$\alpha$ emission) as a
function of {\Large{\it{z}}} - height for all M-dwarf spectral types, suggesting a
decline in the magnetic activity of older M dwarfs.  All results from
this 1D dynamical simulation are consistent with observed activity and velocity trends.

\subsection{Comparison With Previous Studies}

Figure 8 indicates a comparison between the metallicity distribution
(volume-corrected) from this work (solid line) and that from WW12 who
fobtained the metallicity distribution of 4141 M dwarfs in the
spectroscopic SDSS catalog (dashed line). As shown in the Figure,
there is a slight offset between the peaks of the two M-dwarf
distributions. It appears, however, that our sample includes a larger
fraction of metal-poor stars than that of WW12. Moreover, the
distribution of WW12 has a steeper slope towards lower metallicities
than ours.

\begin{figure}[h]
 \epsscale{0.8}
 \plotone{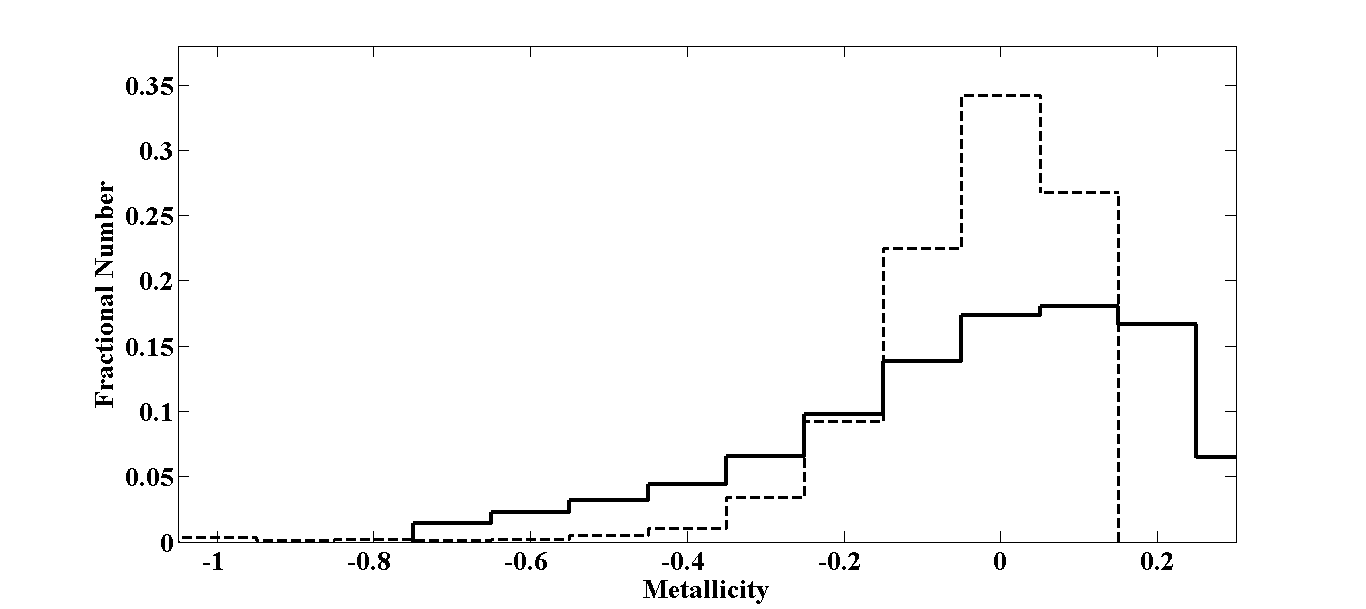}
 \caption{The metallicity distribution of the volume-corrected sample
   from this work (solid histogram) and the distribution from WW12
   (dashed histogram) }
 \end{figure}

Possible reasons for the discrepancy between these two
distributions are as follows. First, the distributions were obtained by
different methods. To estimate the metallicity of their sample, WW12 derived a linear relation (Equation (1) in their paper)
between [Fe/H] and the metallicity-sensitive parameter $\zeta$ defined
in L07. This relation was applicable to M dwarfs with 3500 $\leq$
T$_\text {eff}$ $\leq$ 4000 K and $-1.5 \leq$ [Fe/H] $\leq$ +0.05
dex. Dhital et al.~(2012) recalibrated the definition of $\zeta$ and
showed that this new $\zeta$ could be a significantly better indicator
of metallicity for early M-type dwarfs (between M0 and M3). M13a
tested the parameter $\zeta$ defined in L07 using their own sample and
found that it is not only sensitive to metallicity (and
temperature) but also to some other stellar characteristics such as
activity and surface gravity. This  could lead to an incorrect
identification of some metal-poor stars as near-solar metallicity
stars. They pointed out that although $\zeta$ correlates well with
[Fe/H] for supersolar metallicities, it does not always diagnose
metal-poor M dwarfs correctly and may identify such stars as more metal-rich than they are.  Further, L\'{e}pine et al.~(2013) remarked that
the parameter $\zeta$ defined in L07 overestimated the metallicities of
early-type M dwarfs while the parameter defined in Dhital et
al.~(2012) underestimated the metallicities of these stars, leading them
to the redefinition of the index $\zeta$.  Due to the overestimation
of the metallicity of metal-poor M dwarfs by the parameter $\zeta$
from L07, we could expect a bias toward metal-rich stars in the
metallicity distribution of WW12, as seen in Figure 8.

\begin{figure}[h]
 \epsscale{0.7}
 \plotone{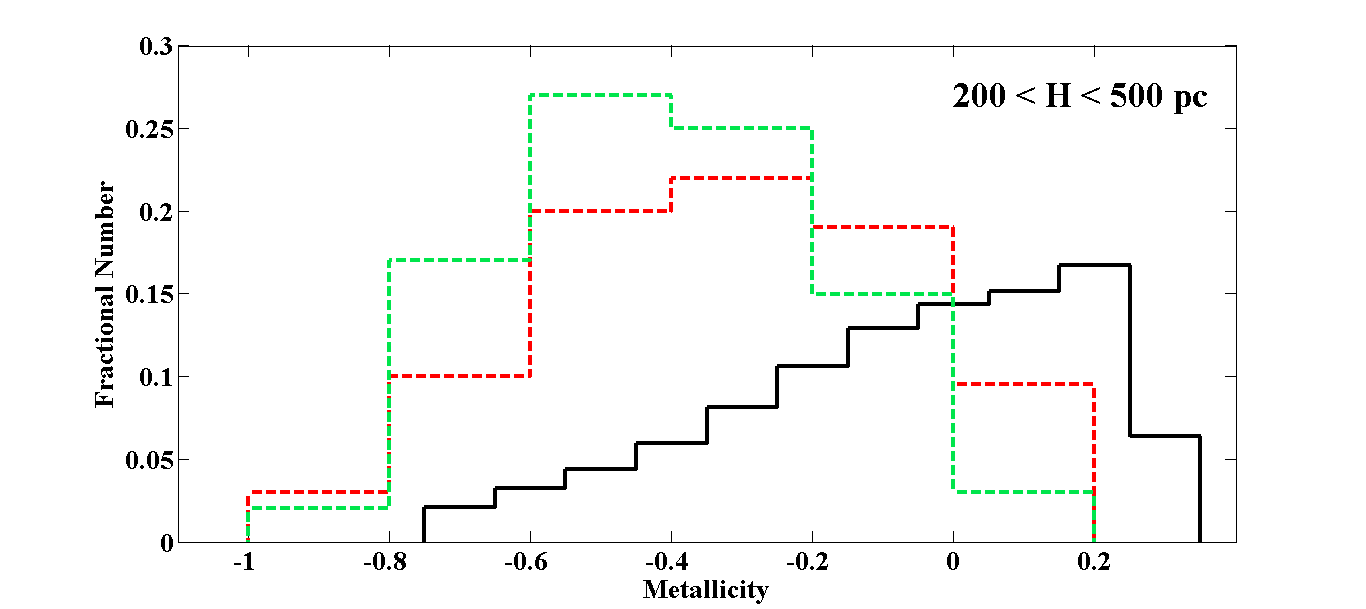}
 \plotone{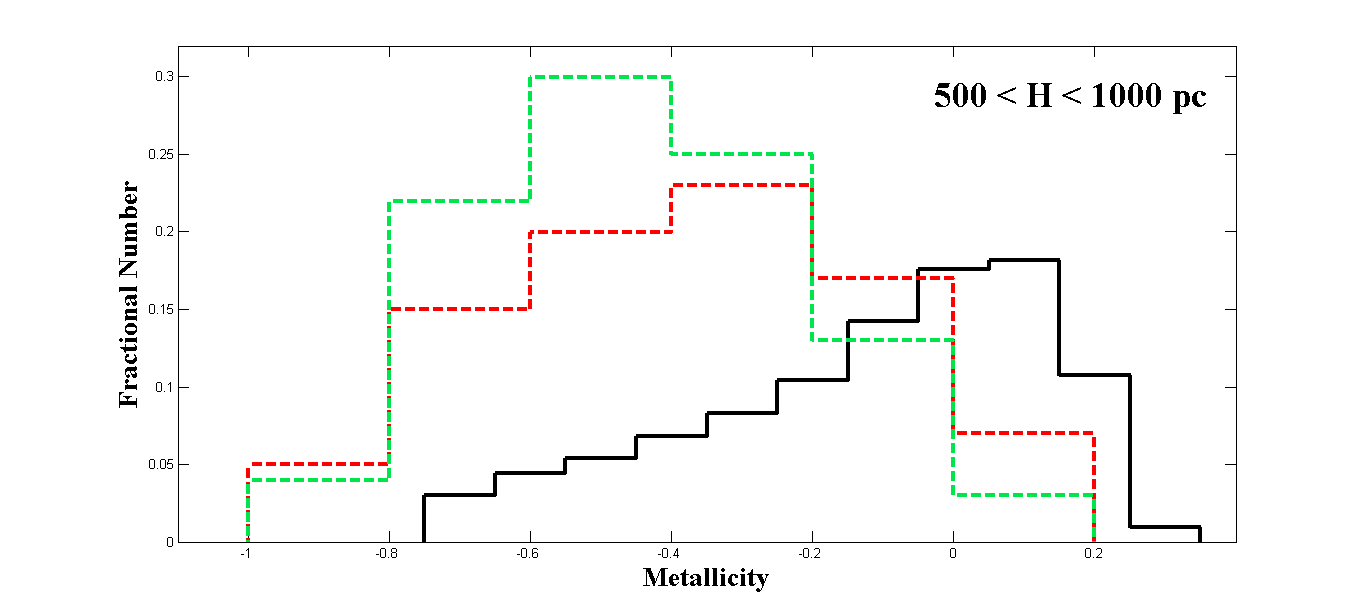}
 \plotone{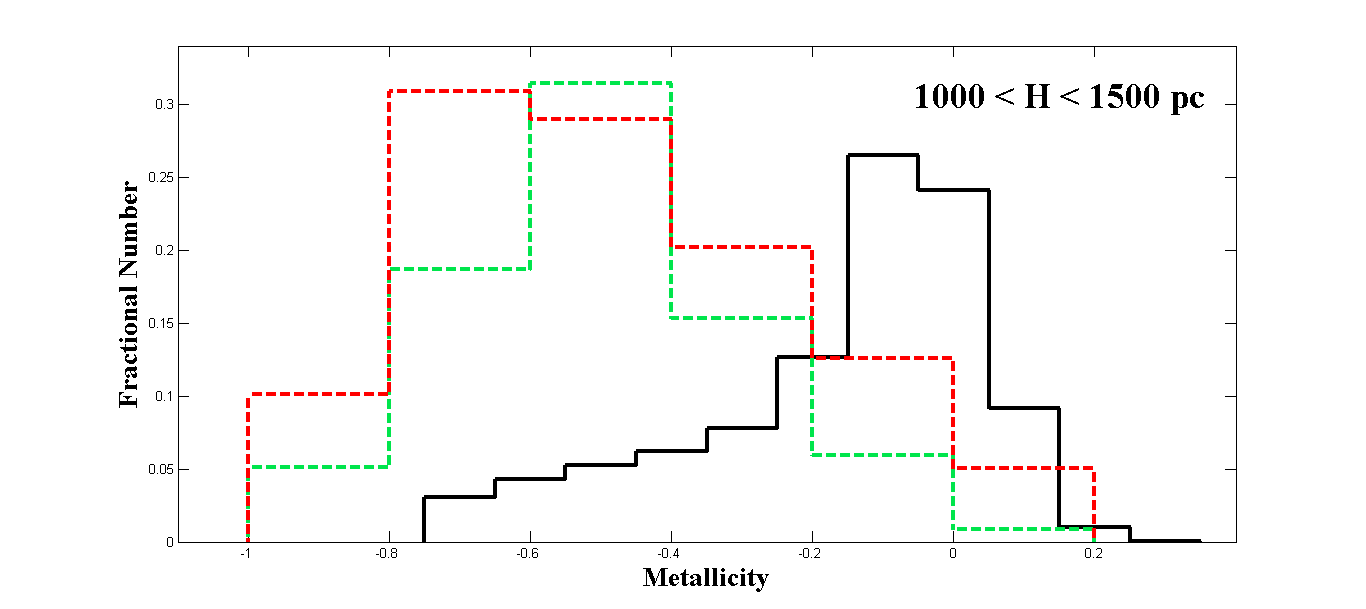}
 \caption{The metallicity distribution of the volume-corrected sample of M dwarfs
   from this work (solid  black histogram) and those of G dwarfs (dashed green histogram) and K dwarfs (dashed red histogram) from S12 for different height ranges. }
 \end{figure}

Second, there is  an important difference between our sample and that of
WW12. As shown in Section 4, our sample is taken mostly from the
Galactic thin disk, and a large fraction of its stars are in the solar
neighborhood with {\Large{\it{z}}} $\lesssim 500$ pc. On the other hand, a
significant portion of stars in the sample of WW12 lies outside the
thin disk and is not representative of the local neighborhood: the WW12's sample is expected to have more low-metallicity stars than ours, but that is not what is seen in Figure 7.

It should be noted that the distribution of WW12 in Figure 8 is 
volume-corrected due to the metallicity-dependent volume
coverage. It should be mentioned that this is different from the volume correction (owing to only volume effects) which we made above. 
WW12 argued that since low-metallicity main sequence stars
(subdwarfs) are less luminous than higher-metallicity stars of the
same temperature or spectral type, higher metallicity stars should be
overrepresented in magnitude-limited samples. In other words,
metal-poor M dwarfs must be, on average, closer than the metal-rich
ones in such samples. Using a sample of stars from Woolf et al.~(2009)
for which parallax data were available, WW12 determined the luminosity
variation with metallicity for M dwarfs in the temperature range 3500
$\leq$ T$_\text {eff}$ $\leq$ 4000 K. Based on this variation, they
calculated volume-correction factors for different values of [Fe/H]
(Table 1 in their paper) and implemented a volume correction for their
metallicity distribution. Accordingly, to have a more reliable
metallicity distribution, we need to determine the
metallicity-dependent correction factors appropriate to our
magnitude-limited sample. Nevertheless, we would not expect these
corrections to change the location of the peak and the slope of the
low-metallicity tail significantly (Figure 2 of WW12). Therefore, the
uncorrected metallicity distribution can give us, to some extent,
important information about the local Galactic disk.

Figure 9 compares our M-dwarf metallicity distributions with those of G and K dwarfs in the SEGUE survey from Schlesinger et al. (2012, hereafter S12) for three different Galactic height ranges. As noted in S12, the G- and K- dwarf metallicity distributions have more metal-poor stars as height increases. It can be seen that the M-dwarf distribution exhibits the same trend, which implies a similar star formation history as G and K dwarfs. There is also a discrepancy between the distributions of different spectral types; the cooler spectral type, the more metal rich the distribution includes.

JA09 argued that the M and FGK dwarfs in the local Galactic thin disk should have the same metallicity distributions and that there should not be a systematic offset between M- and FGK-dwarf metallicity distributions. They related the offset between the metallicity distributions of M dwarfs and FGK dwarfs in some published papers  to the reliability of the methods by which the metallicity of M dwarfs were determined. Accordingly, they adopted the mean M dwarf metallicity of their sample to be the same as the mean metallicity of a volume-limited sample of FGK dwarfs ($\sim -0.05$ dex) from VF05. SL10 improved upon JA09 and remarked that for a fair comparison between samples of M dwarfs and FGK dwarfs in the local Galactic disk, the samples must not only be volume corrected but also have equivalent kinematics. By selecting two volume-completed, kinematically-matched samples of FG and M dwarfs, SL10 derived a mean metallicity of around $-0.14$ dex for M dwarfs in the solar neighborhood. It should be mentioned that we compared our distribution to those of previous works without any consideration of kinematics of the samples under study. To make proper comparisons between the metallicity distributions of dwarf stars with different spectral types, volume-corrected samples with equivalent kinematics have to be considered.

On the other hand, the offset between FGK- and M-dwarf distributions might be explained in other ways. For example, similar to S12, one may expect that since M dwarfs have longer lifetimes than G and K dwarfs, there is a possibility that more metal-rich G and K dwarfs have evolved off the main sequence, causing a metallicity bias in G- and K-dwarf distributions against high-metallicity values which has not happened for cooler M dwarfs. Furthermore, if more lower-mass dwarfs are born than higher-mass stars as the metallicity of interstellar gas increases, there should be fewer metal-poor M dwarfs than G and K dwarfs.

While there have been discrepancies between the metallicity distributions of different spectral types from various studies and some effort has been made to explain them, we can mention the work of C11 whose metallicity distribution of solar type stars (obtained by a photometric method) has a peak around [Fe/H] $\simeq$ 0, close to those of the M-dwarf distributions shown in Figure 8. This is in agreement with the statement of JA09 who pointed out there should not be any offset of different metallicity distributions. To reach an accurate conclusion, more careful investigations are needed.

\section{THE Galactic Chemical Evolution (GCE) MODELS}

\subsection{The Simple Closed Box Model (SCBM)}
The simplest model of GCE, the SCBM, is based on three assumptions:

\begin{enumerate}
 \item The system is isolated and there is no mass inflow or outflow,

 \item There are two types of stars; those that live forever and those
   that die right after their birth, indicating instantaneous
   recycling,

 \item The system initially starts off with entirely metal-free gas
   and eventually ends up full of stars.
 \end{enumerate}

Studies of the chemical evolution of local G dwarfs (e.g., van den
Bergh~1962; Pagel \& Patchett~1975; Wyse \& Gilmore~1995) have shown
the existence of the ``G dwarf problem", which indicates that the SCBM
predicts many more low metallicity G dwarfs than  are
observed. Similarly, the ``K-dwarf problem", a paucity of metal-poor K
dwarfs relative to the prediction of the SCBM, has also been observed
(e.g., Casuso \& Beckman 2004, hereafter CB04). WW12 were the first to
recognize the ``M-dwarf problem" (as shown in Figure 2 of their
paper); the number of low metallicity M dwarfs is insufficient to
match the SCBM, as G and K dwarfs.

To test the SCBM, we compare our metallicity distribution with that
predicted by the model. It
can be shown (e.g., Pagel 2009 hereafter P09; Mo et al.~2010) that the
mass in stars ($\mathrm{M}_\mathrm{S}$) with metallicities between Z
and Z + dZ is given by

\begin{equation}
\text{dM}_\text{S}\ \propto \ \exp(-\frac{\text{Z}}{\text{p}}) \, \text{dZ}
\end{equation}
\vspace{0.12cm}

\noindent
where p is the metal yield, defined as the mass of
newly synthesized and ejected metals per unit mass  permanently 
locked up in the stars. Considering [Fe/H] $\sim$ [M/H] $\sim$
{log}{(}$\frac{\mathrm{Z}}{\mathrm{Z}_{\sun}}${)}, the mass of stars with
metal abundances between [Fe/H] and [Fe/H]+d[Fe/H] can be estimated as

 \begin{equation}
\text{dM}_\text{S}\ \propto \ \text{Z} \exp(-\frac{\text{Z}}{\text{p}}) \, \text{d[Fe/H]},
\end{equation}
\vspace{0.15cm}
 
\noindent
which can be rewritten as
 
\begin{equation}
\frac{\text{dM}_\text{S}}{\text{dlog(Z)}} \ \propto \ \text{Z} \exp(-\frac{\text{Z}}{\text{p}}) 
\end{equation} 
\vspace{0.1cm}
                                                                  
The proportionality (4) is the metallicity distribution function
predicted by the SCBM. The histogram in Figure 10 depicts our
volume-corrected metallicity distribution. The black curve in this
Figure shows the normalized distribution from the SCBM with p=0.025
which has a maximum at the same metallicity ([Fe/H]=+0.1 dex) as our
histogram.

Obviously, the SCBM cannot reproduce the low metallicity tail of the
distribution, suggesting the M dwarf Problem.  It is clear that one or more of the assumptions in
the SCBM, such as instantaneous recycling, the system's isolation
without gas flow or the assumption of a zero initial metallicity are
unrealistic. The G and K dwarf problems have motivated a generation of
more complicated GCE models. We will examine a few such models and
test them using our M-dwarf metallicity distribution in the following
subsections.

\begin{figure}[h]
 \epsscale{0.8}
 \plotone{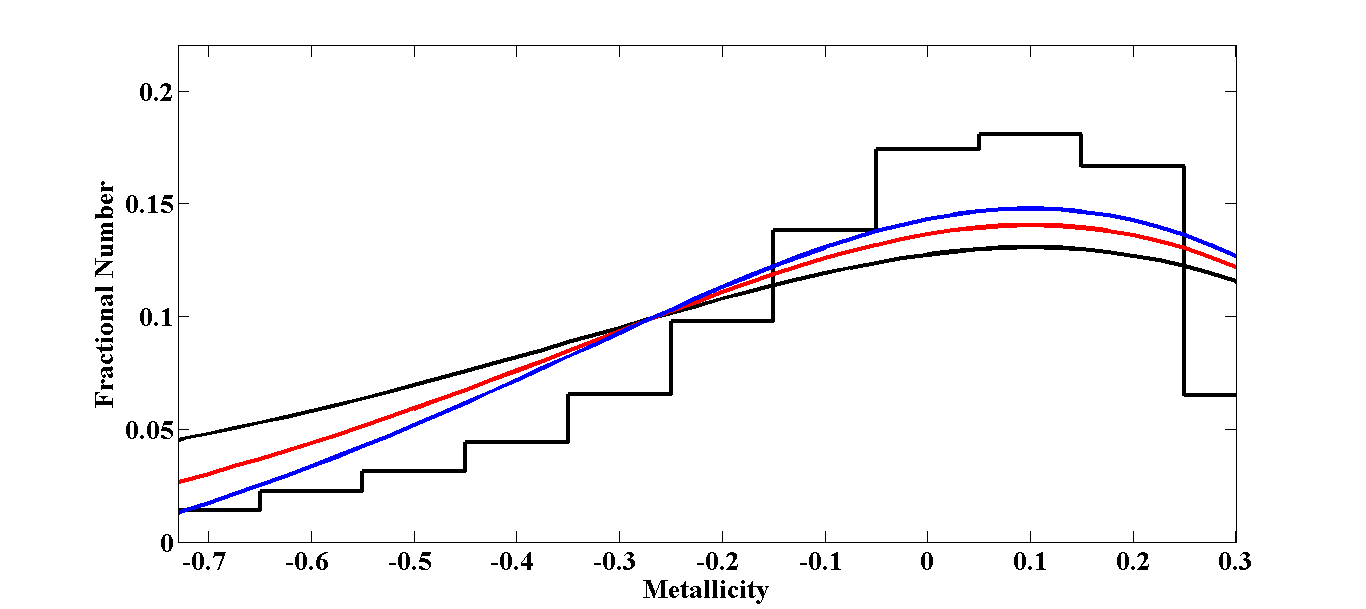}
 \caption{The comparison between the metallicity distribution of M
   dwarfs from this study (the histogram) and those from the SCBM with
   different initial metallicities: $\mathrm{Z}_{0}$ = 0 (the black
   curve), 0.002 (the red curve), and 0.003 (the blue curve). }
 \end{figure}

\subsection{The SCBM With Pre-enrichment}

One of the solutions proposed for the G dwarf problem is to assume an
initial metallicity of $\text{Z}_{0} \neq 0$ (Samland \& Hensler~1996).  If,
for some reason, there is a finite initial abundance of order
[Fe/H]$_{0} \approx -1$ dex, a good fit to the distribution is
obtained at the expense of rejecting some stars with low
metallicity. In this case, all metallicities (Z) in Equation (4) are
replaced by $\text{Z} - \text{Z}_0$ yielding:
                                             
\begin{equation}
\frac{{\mathrm{d}}\text{M}_\text{S}}{\text{dlog(Z)}} \ \propto \ (\text{Z}-\text{Z}_0) \exp(-\frac{(\text{Z}-\text{Z}_0)}{\text{p}}) 
\end{equation} 
\vspace{0.1cm}

The normalized distributions calculated by the SCBM with two selected
initial metal abundances of $\mathrm{Z}_{0}$ = 0.002 (the red curve)
and 0.003 (the blue curve) are shown in Figure 10. We set the yields
for the red and blue curves to p = 0.023 and 0.022, respectively, to
have a maximum at [Fe/H]=+0.1 (the same as our distribution). As can
be seen, the distributions of the SCBM with a pre-enrichment give a
better fit to our measured metallicity distribution than that without
any initial abundances (the black curve).

There have been several hypotheses to justify such a
pre-enrichment. For example, this might arise from prior star formation
activity in the halo (Ostriker \& Thuan~1975) or the bulge. It
was argued (Prantzos~2007), however, that while the Galactic halo reached a
maximum metallicity of Z $\sim$ 0.002 ($\sim$ 0.1 $\mathrm Z_{\sun}$),
its mean metallicity is Z $\sim$ 0.0006 ($\sim$ 0.03 $\mathrm
Z_{\sun}$) and its total mass (2$\times 10^{9}$ $\mathrm M_{\sun}$) is
20 times less than that of the disk (4.5$\times10^{10}$ $\mathrm
M_{\sun}$).  K\"{o}ppen \& Arimoto~(1990) suggested a
model in which the bulge evolves with a large yield (p $\approx$
0.034) and ends star formation by ejecting 1/10 of its mass as highly
enriched gas with an abundance of 0.05 ($\sim$ 2.5 $\mathrm Z_{\sun}$)
in a terminal wind which is captured by the proto-disk. Assuming that the
proto-disk has a mass equal to that of the bulge, this mass transfer
results in an initial disk abundance of 0.005 ($\sim$ 0.25 $\mathrm
Z_{\sun}$) after mixing. Such a model is consistent with the metallicity
distribution functions of both the bulge and disk (for more
discussions, see P09).
   
On the other hand, the p values derived from these
models above are larger than the estimated value in the solar
neighborhood: $\mathrm p_{\mathrm {SolN}}$ $\sim$ 0.7 $\mathrm
  Z_{\sun}$ = 0.014 (P09). We therefore need to establish models
  which not only match the metallicity distributions of the solar neighborhood but also give yields more consistent with observational
  data.
   
\subsection{Inflow Models With Declining Rates} 
In more realistic models, the disk does not evolve as a closed box and
gas inflow or outflow (or both) is present. Inflow models have been
attractive because of their capacity to reduce the
number of low-metallicity stars born at early times, providing an
elegant solution of the G dwarf problem, especially in view of the
fact that  gas accretion is expected to be a common phenomenon in
the Universe (e.g., Prantzos~2007). There is indirect observational
evidence for gas infall into the Galactic disk. The star-formation
rate in the disk is about a few solar masses per year, while the total
gas mass therein is $\sim 5\times10^9$ $\mathrm M_{\sun}$. Thus, if there were
no accretion of new material, our disk would run out of gas in a
fraction of the Hubble time (Mo et al.~2010).

The inflowing gas could originate from the gaseous halo, the bulge,
the outer parts of the disk, or the accretion of small satellite
galaxies. The disk itself may be considered as having formed completely
from such inflowing gas or as having an initial mass (usually small
compared to its final mass) with or without metals. Still more
physically realistic models assume gas inflow with variable rates, and
those with declining rates appear especially to provide in better agreement
with recent observations (P09). In this section, two  inflow
models will be considered and the resulting metallicity distributions
will be compared with that from this study.

\subsubsection{The Exponential Inflow Model (EIM)}
Some studies have shown that an exponentially decreasing infall rate
with a long characteristic time scale $\sim$ 7 Gyr can provide a
reasonably good fit to the data. For example, CB04 demonstrated that such a
model  seems to reproduced
the observed metallicity distributions of G and K dwarfs in the
 disk.

For simplicity, some inflow models are based on a
specific star formation rate given by

\begin{equation}
\frac{\text{dM}_\text{S}}{\text{dt}} = \text{\it{w}}\,{\mathrm{M}_\mathrm{G}}(\mathrm{t})
\end{equation} 
\vspace{0.1cm} 

\noindent
where t is time, ${\mathrm{M}_\mathrm{G}}$(t) is the gas mass  as a function of time  and $w$ is a constant. By defining a time-like variable u by
\begin{equation}
{\mathrm{u}} = w{\mathrm{t}},
\end{equation}
\vspace{0.1cm}
Sommer-Larsen (1991) constructed a dynamical model with the assumption

\begin{equation}
{\mathrm{F}}({\mathrm{u}})\ =\ A\, w\, \exp({-{\mathrm{u}}})
\end{equation}
\vspace{0.1cm} 

\noindent 
in which F is the accretion rate of material from outside the system
and A is a constant. If the initial mass and metallicity of the system
are assumed to be zero and if the inflow gas has no metal content
($\mathrm Z_{\mathrm F}$ = 0) then, as u becomes very large,
 the expression for Z and
${\mathrm{dM}_\mathrm{S}}$/dlog(Z) can simply be written as (P09):

\begin{equation}
{\mathrm{Z}}({\mathrm{u}}) =   \big(\frac{\text{pu}}{\text{2}})
\end{equation}
\vspace{0.05cm}

\begin{equation}
\frac{{\mathrm{d}}\text{M}_\text{S}}{\text{dlog(Z)}} \ \propto \ \big(\frac{\text{Z}}{\text{p}}\big)^2 \ \exp(-2\text{Z}/{\text{p}}) 
\end{equation}
\vspace{0.15cm}  

The expression on the right side of the proportionality (10) is
essentially the square of that corresponding to the SCBM. The
normalized distribution from the EIM for p = 0.025, which peaks at
[Fe/H] = +0.1, is shown in Figure 11 (the red curve); it
presents the metallicity distributions from this work (the
histogram) and from the SCBM (the black curve) as well. Evidently, the
EIM predicts fewer metal-poor M dwarfs and a steeper slope in the low
metallicity tail than the SCBM, and is therefore in better agreement with
the  data. There is, however, an inconsistency between the value
of p from this model and $\mathrm P_{\mathrm {SolN}}$, which is
  somewhat problematic. It should be remarked here that if our
  distribution had a peak at a lower metallicity, say, [Fe/H] $\approx
  -0.1$ dex, the model would result in a more reasonable
  yield. Accordingly, the accuracy of the metallicity distribution is
  critical for a comparison of GCE models.

\begin{figure}[h]
\epsscale{0.8}
\plotone{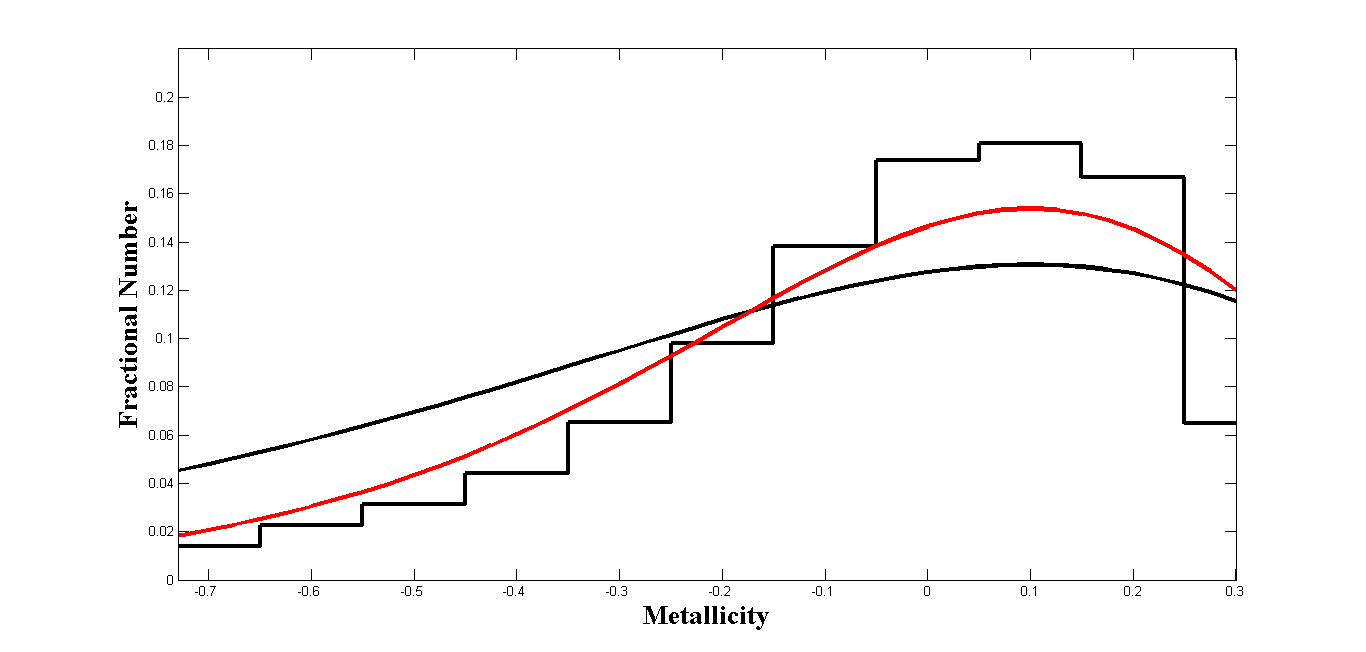}
\caption{The comparison between the metallicity distributions of M
  dwarfs from this study (the histogram), the EIM (the red curve) and
  the SCBM (the black curve).}
\end{figure}
 
\subsubsection{Clayton's Models}

Clayton (1985; 1987; 1988) has derived a series of models in which the
inflow rate is parameterized by

\begin{equation}
\text{F(t)} = \frac{\text{k}}{\text{t + }{\mathrm{t}}_0} \, {{\mathrm{M}_\mathrm{G}}(\mathrm{t})}
\end{equation} 
\vspace{0.1cm} 

\noindent
where k is an integer and $\mathrm{t}_{0}$ (or $\mathrm{u}_{0} = w
\mathrm{t}_{0}$) is arbitrary. By considering Equations (6) and (7),
and assuming ${\mathrm{Z}_\mathrm{0}} = \mathrm{Z}_\mathrm{F} = 0$,
 expressions for Z and
$\mathrm{\mathrm{dM}_\mathrm{S}}/\mathrm{dlog(Z)}$ in terms of u can
be derived (Equations (8.38) and (8.39) in P09). The red curve in
Figure 12 shows a distribution of this kind with k = 7 and p = 0.017,
having a peak at [Fe/H] = +0.1 dex. The P value in this case is in
better agreement with $\mathrm P_{\mathrm {SolN}}$. Although the model
fits the low-metallicity tail of our distribution (the histogram) well
compared with the SCBM (the black curve), it underestimates the number
of supersolar-metallicity stars, and a modification of this model for
the high-metallicity tail is needed.

\begin{figure}[h]
\epsscale{0.8}
\plotone{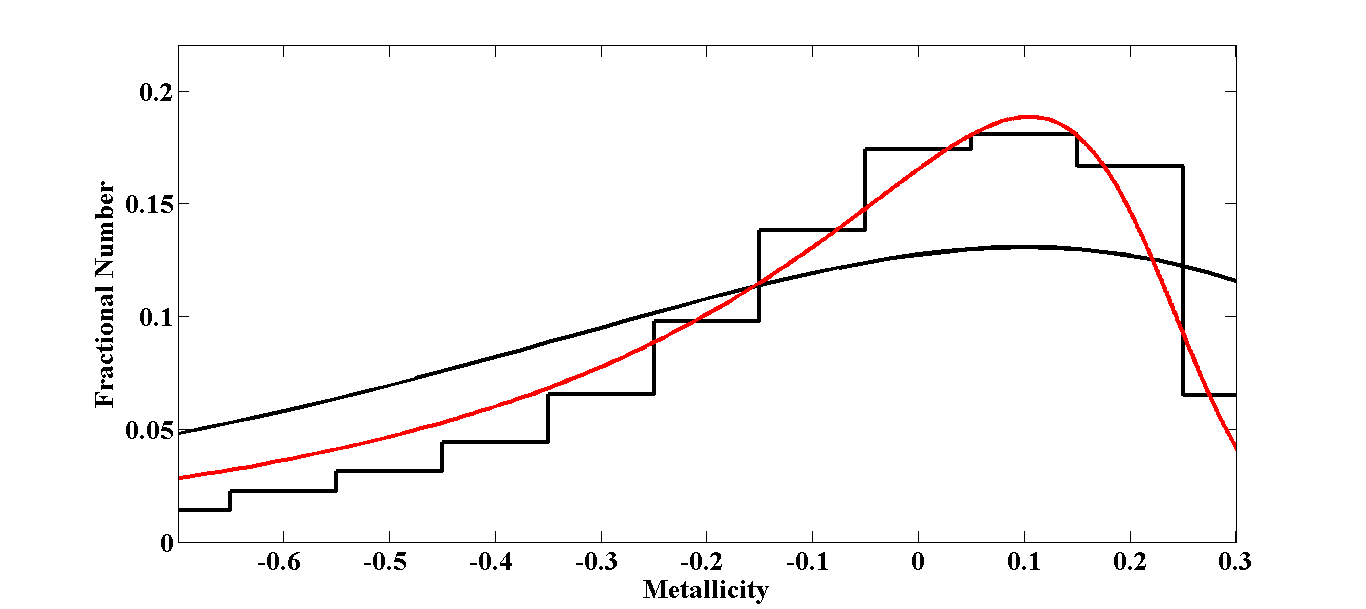}
\caption{The comparison between the metallicity distributions of M
  dwarfs from this study (the black histogram), the Clayton's model (the red
  curve) and the SCBM (the black curve).}
\end{figure}

There have been  other  models such as merger models (Nagashima \& Okamoto 2006)  which resolve the G
and K dwarf problems and  can be tested using M-dwarf metallicity
distributions in future. However, larger, more accurate
data sets are required in order to discriminate among these models and find the more realistic one.
  
The metallicity distribution of long-lived stars (i.e., G, K and M
dwarfs) as well as the age-metallicity relation traced by the
metallicity of these dwarfs are the most important observational
constraints for GCE. These low-mass stars belong to stellar populations
of different ages and their metallicity distribution provides a fairly
complete record of the evolutionary history of the Galaxy. However,
CB04 pointed out that G dwarfs are sufficiently massive that they have
begun to evolve away from the main sequence, and consequently, K
dwarfs would make a cleaner sample for the local metal abundance
distribution.  For the same reason, we can expect that M dwarfs
 would provide an even better
sample to represent the local metallicity distribution.  M dwarfs are
 numerous, making up around half the total stellar mass of
the Galaxy (RG97), 
and their metallicity distribution therefore
offers a robust tool for studying the chemical evolution of the Milky Way.  However, more precise metallicity calibrations are essential.

\section{SUMMARY}  

Using a sample of 67 carefully selected dwarfs, we developed an
optical-NIR photometric method to determine the metallicity of dwarf
stars with spectral types between K6 to M6.5 in the metallicity range
$-0.73$ to +0.3 dex. The calibration sample has two parts; the first
part includes 18 M dwarfs in common proper pairs with an FGK star or
early-type M dwarf of known metallicity and the second part contains
49 dwarfs with metallicities obtained through an analysis of
moderate-resolution spectra. Although the method may not be as
accurate as moderate-to-high resolution spectroscopic techniques, they
can be consistently applied to  large numbers of stars, facilitating
statistical investigations on metallicity distributions efficiently.

We selected a large sample of around 1.3 million M dwarfs from the merged
SDSS and 2MASS catalogs and corrected their {\it gJK} photometry for
Galactic extinction based on a widely used dust map of the northern
Galactic hemisphere. These stars meet the requirements for clean
photometry and fall in color ranges necessary for removing possible
giants as well as those ranges required for our metallicity
calibration. We also applied a color cut needed for a photometric
parallax to estimate stellar distances. Using the {\Large{\it{z}}} - height
distribution, we found that the majority of M dwarfs in this large
sample are found in the Galactic thin disk. By applying our
calibration to these M dwarfs, the metallicity distribution of the
local thin disk was determined and investigated. The metallicity
distributions of subsamples with different {\Large{\it{z}}} - heights
recovers the well known observation that there is a decrease in mean
metallicity with {\Large{\it{z}}} - height, confirming yet again that older stars have on
average lower metallicity and are farther from the Galactic plane.

We examined the SCBM of GCE by comparing the distribution expected
from this model with our M-dwarf distribution. A significant
discrepancy between the model and our results  in the low-metallicity
tail leads to the ``M dwarf problem,'' similar to the previously known
G and K dwarf problems. More realistic models
advanced to resolve the G and K dwarf problems  may also be
 solutions to the M dwarf problem as well. We explored a few
of these models such as the SCBM with pre-enrichment and two kinds of
infall gas models with declining rates, and showed that these could
more or less mitigate the M dwarf problem.

In order to improve significantly these results and to determine a
more reliable M-dwarf metallicity distribution, a more accurate
calibration over a broader range of metallicities is required.  The
photometry of many M dwarfs of known metallicity is unusable because
the images are saturated in SDSS images which accounts for their
exclusion from our calibration sample and why only the $g$ band was
employed in the optical region.  The acquisition of high-quality
photometric $griz$ data for a sample of M dwarfs with a variety of
well-determined metallicities  is already in
progress.

\section*{ACKNOWLEDGMENTS}
 We would like to thank the anonymous referee for many thoughtful
 suggestions that improved our manuscript.  We are also indebted to
 Patrick Hall for his helpful discussions involving the M-dwarf
 calibration sample and for his guidance in the statistical analysis
 of the large sample. We kindly thank Andrew Mann, John Bochanski,
 Andrew West, S\'{e}bastian L\'{e}pine, Ryan Terrien, Vincent Woolf,
 Rohit Deshpande, Chad Bender, and Zeljko Ivezi\'{c} for their helpful
 suggestions.  We are grateful for assistance provided by Xiaoyi Dong
 while the work was being undertaken.
 
Funding for SDSS-III has been provided by the Alfred P. Sloan
Foundation, the Participating Institutions, the National Science
Foundation, and the U.S. Department of Energy Office of Science. The
SDSS-III web site is http$://$www.sdss3.org$/$. SDSS-III is managed by
the Astrophysical Research Consortium for the Participating
Institutions of the SDSS-III Collaboration including the University of
Arizona, the Brazilian Participation Group, Brookhaven National
Laboratory, Carnegie Mellon University, University of Florida, the
French Participation Group, the German Participation Group, Harvard
University, the Instituto de Astrofisica de Canarias, the Michigan
State$/$Notre Dame$/$JINA Participation Group, Johns Hopkins
University, Lawrence Berkeley National Laboratory, Max Planck
Institute for Astrophysics, Max Planck Institute for Extraterrestrial
Physics, New Mexico State University, New York University, Ohio State
University, Pennsylvania State University, University of Portsmouth,
Princeton University, the Spanish Participation Group, University of
Tokyo, University of Utah, Vanderbilt University, University of
Virginia, University of Washington, and Yale University.

 This research has made use of data products from the Two Micron All
 Sky Survey, which is a joint project of the University of
 Massachusetts and the Infrared Processing and Analysis
 Center/California Institute of Technology, funded by the National
 Aeronautics and Space Administration and the National Science
 Foundation.

MMDR gratefully acknowledges the support of Natural Sciences and
Engineering Research Council in carrying out this research.

\begin{deluxetable}{rrrrrrrrrrrr}
\tabletypesize{\footnotesize}
 \rotate
 \placetable{t}
 \tablecolumns{5}
 \tablewidth{0pc}

 \tablecaption{Observational Properties of 18 M dwarfs with metallicities determined by high-resolution spectroscopy}
 \tablehead{
 \colhead {Name} & \colhead {RA (deg)\tablenotemark{\it{a}}} & \colhead {DEC (deg)\tablenotemark{\it{a}}} &  \colhead {\it g}& \colhead {{\it g}.stat.err\tablenotemark{\it{b}}} & \colhead {{\it g}-Satur\tablenotemark{\it{c}}} &  \colhead {\it J} & \colhead {{\it J}.tot.err\tablenotemark{\it{d}}}& \colhead {\it H} & \colhead {{\it H}.tot.err\tablenotemark{\it{d}}} &  \colhead {\it K} & \colhead {{\it K}.tot.err\tablenotemark{\it{d}}}}
 
 \startdata 
 
LJ1210+1858E &   182.5412&	18.9690 & 21.234  &0.059 & un.sat& 13.691& 0.027	&13.052& 0.033	&12.692 &0.024\\
LJ1000+3155 & 150.2096&	31.9294& 17.352 &0.039&	w.sat& 	10.261	& 0.018& 9.643& 0.016 &	9.275& 0.018\\
LJ1248+1204&	192.2227	&12.0757	&16.871& 0.028	&un.sat	&11.400&0.021	&10.871&0.024 	&10.570	&0.023	 \\						
LJ1604+3909W&	241.2122&	39.1600&	15.122&0.015 &	un.sat&	9.903&	0.021 &9.453&0.021	&	9.159&0.017	\\						
LJ1425+2035W &	216.3579&	20.5960&	17.898&0.240	&w.sat	&12.462&0.029	&	12.006&	0.030 &11.716&0.023 \\							
2M 1743+2136& 	265.8142&	21.6028&	17.867&0.213 &	w.sat&	11.511&0.025&	11.016&	0.021	&10.700	&0.019	 \\						
GJ 3628 B	&162.65939&	51.75045&	14.931&0.019 &	un.sat&	9.828&0.022&	9.247&0.021&	9.015	&0.018 \\						
NLTT 40692 &	233.8569	&60.0855	&14.311&0.019 &	un.sat&	9.270&	0.022&8.700&	0.021&8.412	&0.018	 \\	
LJ0849+0329W&	132.2594&	3.4964	&15.997&0.018 &	un.sat&	10.756&0.024&	10.176&	0.021&9.911	&0.023 \\			
NLTT 2478	&11.3066&	0.2642&	15.323&0.024&	un.sat&	10.114&0.027&	9.561& 0.023	&	9.264	&0.021 \\						
NLTT 57675 &	355.4381	&$-5.9707$	&15.314&0.027 &	un.sat&	10.392&	0.024 &9.809&0.021	&	9.583	&0.021\\ 
NLTT 42396& 	243.7207	&60.6411&	14.680&0.017 &	un.sat&	9.818&	0.020	&9.295&	0.019	&9.023	&0.014	\\						
Gl173.1B	&69.9303&	9.8630&	15.056&0.013&	un.sat&	10.263&0.022 &	9.715&	0.028&9.421	& 0.024	\\						
NLTT 39578 &	227.96436&	39.5507&	14.409&0.013 &	un.sat&	9.873&0.021&	9.276&0.016 &	9.069&0.018	\\							
LJ1237+3549 	&189.3145&	35.8216&	15.511&0.024	&	un.sat&	11.353&0.019&	10.757&	0.017 &10.519& 0.019	\\							
NLTT 36190	&211.2326&	1.9564&	14.330&0.016& 	un.sat&	10.130&0.026	&9.483&0.023 &	9.269&0.019	\\							
NLTT 28180  &	175.0868	&9.5126&	14.252&0.018&	un.sat&	10.115&0.023 &	9.544&0.025 &	9.309 &0.021	\\							
LHS 3084	&233.66708&	2.20412&	14.279&0.017 &	un.sat	&10.532&0.023 &	9.989&	0.025&9.783&0.023\\

 \enddata 
 
  \tablenotetext{\it{a}}{\small{The astrometry is taken from 2MASS observations.}}
  \tablenotetext{\it{b}}{The statistical photometric uncertainty in the {\it {g}}-band photometry, the systematic errors ($\sim 2\%$, Ivezi\'{c} et al.~2004 ) are not included.}
  \tablenotetext{\it{c}}{The saturation state of {\it{g}} band, un.sat = unsaturated and  w.sat = weakly saturated}
 \tablenotetext{\it{d}}{The total photometric uncertainty, including the corrected band photometric uncertainty, the nightly photometric zero-point uncertainty and the flat-fielding residual error}
 \tablecomments{\small {Name abbreviation: LJ=LSPM J}}
 \tablecomments{\small{ In all uncertainties above, the errors due to extinction corrections are not considered.}}
  \end{deluxetable}

\begin{deluxetable}{rrrrrrrr}
\tabletypesize{\footnotesize}
\rotate
 \placetable{t}
 \tablecolumns{5}
 \tablewidth{0pc}
 \tablecaption{Measured Properties of 18 M dwarfs with metallicities determined by high-resolution spectroscopy}
 \tablehead{
 \colhead {Name} & \colhead {Sp.Type} & \colhead {Sp.Type Ref.} & \colhead {Primary}&  \colhead {Sp.Type$_\text {Prim}$}& \colhead {Sp.Type$_\text {Prim}${Ref.}}& \colhead {[Fe/H]$_\text {Prim}$} & \colhead {[Fe/H]$_\text {Prim}${Ref.}} }
\startdata 
 LJ1210+1858E&	M6.5&	 M14&HIP 59310&	K3	&M14
&	+0.30 $\pm$ 0.03&	 M13a-Calib.Sample\\
LJ1000+3155&	M6&	 M13a& HIP 49081 &	G3	&Simbad
	&+0.20 $\pm$ 0.03&	 VF05\\
LJ1248+1204&	M5	& N14& HD 111398&	G5&	Simbad
	&+0.08 $\pm$ 0.03	& VF05\\
LJ1604+3909W&	M5	& N14& HD 144579	&G8&	Simbad
&	$-0.69$ $\pm$ 0.03&	 VF05\\
LJ1425+2035W &	M5&	 M14&HIP 70520	&F9	&M14
	&$-0.57$ $\pm$ 0.05&	 Ram07\\
2M 1743+2136& 	M4.5	& M14&HIP 86722&	K0	&M14&
	$-0.39$ $\pm$ 0.05&	 F08\\
GJ 3628 B&	M4.1	& M13a	& HIP 53008	&G5	&Simbad
&$-0.04$ $\pm$ 0.10	& SOP\\
NLTT 40692 &	M4.1	& M13a& HIP 76315 &	K3	&Simbad
	&+0.11 $\pm$ 0.03	& VF05\\
LJ0849+0329W&	M4&	 N14&	HD 75302&	G5&	Simbad
&+0.10 $\pm$ 0.03&	 VF05\\
NLTT 2478&	M3.8&	 M13a&HIP 3540&	F8&	Simbad
	&+0.02 $\pm$ 0.03	& VF05\\
NLTT 57675 &	M3.6&	M13a&HIP 116906	&G5&	Simbad
	&$-0.03$ $\pm$ 0.03&	 VF05\\
NLTT 42396 &	M3.5&	 M13a&HIP 79629&	G5	&Simbad
	&$-0.25$ $\pm$ 0.04	& M13a-Calib.Sample\\
Gl173.1B&	M3.2	& M13a& HIP 21710&	K2	&Simbad
	&$-0.34$ $\pm$ 0.03	& N12\\
NLTT 39578 &	M2.8&	 M13a&	HIP 74396 &	G5	&Simbad
&$-0.09$ $\pm$ 0.03&	 M13a-Calib.Sample\\
LJ1237+3549 &	M1.8	& M13a& 	HIP 61589&	G0	&Simbad
&$-0.05$ $\pm$ 0.03	& M13a-Calib.Sample\\
NLTT 36190&	M1.8	& M13a&	HIP 68799&	K0	&Simbad
&$-0.03$ $\pm$ 0.03&	 M13a-Calib.Sample\\
NLTT 28180  &	M1.6& M13a&HIP 56930&	K0&	Simbad
	& $-0.12$ $\pm$ 0.03 &	 M13a-Calib.Sample\\
LHS 3084 &	M0 &	 WW05/R10-13 & $-$	& $-$  & $-$ &$-0.73$ $\pm$ 0.05\tablenotemark{\it{a}}	& WW05 \\
\enddata 
 \tablenotetext{\it{a}}{\small{This metallicity was determined by direct measurements on the high-resolution spectrum of LHS 3084.}} 
\tablecomments{\small {\textbf{Source of Spectral types:} M13a=Mann et al.~(2013a); M14=Mann et al.~(2014); N14=Newton et al.~(2014);Simbad=SIMBAD  Astronomical Database; WW05/R10$-$13 = Temperature taken from WW05 and spectral type estimated by the temperature-spectral relations of Rajpurohit (2010, hereafter R10; 2013, hereafter R13)
\textbf{Source of [Fe/H]:} 
VF05=Valenti \& Fischer (2005) using the software package SME (Spectroscopy Made Easy analysis, Valenti \& Piskunov 1996) with an adopted uncertainty of 0.03 dex for [Fe/H] and [M/H] through high-resolution observed spectra; SOP=Bouchy \& The Sophie Team (2006), Metallicity taken from spectra of the SOPHIE Spectrograph through high-resolution observed spectra;
N12=Neves et al.~(2012), based on the method of Santos et al.~(2002;2004) through high-resolution observed spectra;
F08=Fuhrmann (2008) through high-resolution observed spectra;
Ram07=Ram${\acute{\mathrm{i}}}$rez et al.~(2007) through high-resolution observed spectra;
M13a-Calib.Sample=Mann et al.~(2013a) in their calibration sample, based on the software package SME as mentioned above and a set of tuned lines from the SPOCS catalog (Valenti \& Fischer 2005) through high-resolution observed spectra; WW05=Woolf and Wallerstein (2005) through high-resolution observed spectra
}}
 
\end{deluxetable}

\begin{deluxetable}{rrrrrrrrrrrr}
\tabletypesize{\scriptsize}
\rotate
 \placetable{t}
 \tablecolumns{5}
 \tablewidth{0pc}
 \tablecaption{Astrometry\tablenotemark{\it{a}} and Extinction-Corrected Photometry of 49 dwarf stars in the calibration sample}
\tablehead{
 \colhead {Name} & \colhead {RA (deg)} & \colhead {DEC (deg)} &  \colhead {\it g}& \colhead {{\it g}.stat.err\tablenotemark{\it{b}}} & \colhead {{\it g}-Satur\tablenotemark{\it{c}}} &  \colhead {\it J} & \colhead {{\it J}.tot.err\tablenotemark{\it{d}}}& \colhead {\it H} & \colhead {{\it H}.tot.err\tablenotemark{\it{d}}} &  \colhead {\it K} & \colhead {{\it K}.tot.err\tablenotemark{\it{d}}}}
 \startdata 

LJ0856+1239	&134.0815	&12.6639&	15.269&0.028	&un.sat&	9.555&0.021&	9.014&0.022&	8.696&0.017\\
LJ1005+1703 &	151.3184&	17.057&	16.786&0.024	&un.sat&	11.106&0.021	&10.547	&0.021&10.246&0.016\\
LJ1530+0926 &	232.6264	&9.4337&	15.941&0.014&	un.sat	&9.534&0.025&	8.984&0.025	&8.647&0.025\\
LJ2012+0112&	303.2498&	1.2162	&16.062&0.038&	un.sat&	10.363&0.026&	9.809&0.024&	9.535&0.019\\
LJ0738+1829 &	114.7117&	18.4890&	17.023&0.012&	un.sat	&10.740	&0.022&10.115&0.023&	9.7861&0.018\\
LJ1002+4827&	150.7057&	48.4593	&16.293&0.024&	un.sat&	9.953&0.019&	9.327&0.016&	9.001&0.015\\
LJ1432+0811&	218.0354&	8.1920	&16.451&0.015&	un.sat&	10.081&	0.024&9.512&0.022&9.155&0.021\\					
LJ1631+4051&	247.8283&	40.8643	&15.620&0.014	&un.sat&	9.427&	0.023&8.847&0.023&	8.492&0.016\\			
LJ0738+4925 &	114.6967	&49.4242	&16.754&0.014	&un.sat&	10.571&0.024	&9.957&0.030	&9.676&0.021\\		
LJ1239+0410	&189.9447&	4.1798	&17.602&0.023&	un.sat	&11.036&0.023	&10.417&0.023&	10.030&0.019\\					
LJ1309+2859	&197.3956&	28.9852	&14.993&0.023&	un.sat	&9.455&0.027	&8.899&0.031	&8.604&0.019\\		
LJ1021+0804& 	155.3295&	8.0741&	16.332&0.019&	un.sat	&10.742&0.022&	10.203&0.022&	9.915&0.021\\		
LJ1352+6649&	208.2108&	66.8185&	16.587&0.023&	un.sat	&10.532&0.020&	9.950&0.015&	9.644&0.019\\			
LJ0918+6037W	&139.5959&	60.6253&	16.620&0.020	&un.sat&	10.983&0.025&	10.379&0.022	&10.093&0.019\\		
LJ0001+0659 &	0.3158&	6.9932&	17.492&	0.018&un.sat&	11.274&0.022&	10.733&0.028&	10.413&0.021\\				
LJ0237+0021	&39.3738&	0.3576	&15.848&0.019&	un.sat	&10.513&0.026&	9.945&0.022&	9.675&0.021\\		
LJ0917+5825	&139.4417	&58.4229&	16.050&0.031&	un.sat	&10.243	&0.022&9.686&0.023&	9.395& 0.020\\					
LJ1031+5705 &	157.8782&	57.0883	&15.472&0.015&	un.sat	&9.716&0.024&	9.176&0.028&	8.876&0.020\\				
LJ1112+0338&	168.1610&	3.6460&	16.734&0.020&	un.sat	&11.070&0.024&	10.533&0.022&	10.232&0.019\\				
LJ1419+0254	&214.8733&	2.9101	&15.732&0.014&	un.sat	&9.932&0.024	&9.3477&0.021&	9.063&0.021\\
LJ0024+2626  &	6.0157	&26.4417&	15.322&	0.017&un.sat	&10.171&0.020	&9.559&0.019&	9.278&0.017\\
LJ1348+0429& 	207.0382	&4.4880&	16.544&	0.021&un.sat&	10.728&0.027&	10.164&	0.022&9.832&0.019\\	
LJ1101+0300&	165.3319&	3.0048&	14.821&0.015&	un.sat	&9.6744&0.027&9.209&0.026&	8.893&0.021\\			
NLTT 8870&	41.4216&	44.9509&	16.813&	0.011& w.sat&11.012&0.021&	10.490&	0.023&10.154&0.018  \\			
LJ0959+4712&	149.9415	&47.2032	&14.903&0.018&	un.sat&	9.750&0.025&	9.195&0.021	&8.920&0.016\\
LJ1148+5305	&177.1970	&53.0860	&15.376&0.023&	un.sat&	10.118&0.023&	9.548&0.027&	9.278&0.024\\
LJ1110+4757&	167.7146&	47.9506&	15.292&0.020&	un.sat&	10.068&0.024&	9.524&0.032&9.265&0.020\\									
LJ0958+0558 	&149.7354&	5.9667	&15.093&0.015&	un.sat&	9.817&0.023&	9.317&	0.027&8.990&0.021\\									
LJ1316+2752	&199.1368	&27.8749&	14.071&	0.022&un.sat&	9.249&0.019&	8.717&0.029&8.437&0.020\\
LJ1709+3909 &	257.3584&	39.1607	&14.640&0.014&	un.sat	&9.819&0.022&	9.276&0.020&	9.025&0.019\\	
LJ1240+1946	&190.0783&	19.7699&	15.301&0.019&	un.sat&10.534&0.026&	9.936&0.032&	9.706&0.021\\ 					
LJ0920+0322&	140.2414&	3.3684&	13.938&0.021&un.sat&	9.306&0.023&	8.761&	0.046&8.497&0.025\\	
LJ1345+2852&	206.2960	&28.8670	&14.483&0.018	&un.sat	&9.862&0.022&	9.298&0.021	&9.046&0.016\\				
LP 655$-$23&	67.7168&	$-8.8220$	&15.055&0.023	&un.sat&	9.824&0.024&	9.266&0.022&	8.978&0.019\\
LJ1314+4846&	198.6227&	48.7780	&13.919&0.019&	un.sat&	9.619&0.022&	9.039&0.028&	8.807&0.022\\									
LJ1038+4831  &	159.6242&	48.5291	&14.274&0.020&	un.sat&	9.485&0.021&	8.832& 0.021	&8.583& 0.017\\		
LJ2148+0126	&327.0426&	1.4451&	14.456&0.016&un.sat&	9.763&	0.024&9.151&0.027&	8.900&0.021\\						
KIC 5252367	&283.1939	&40.4956&	16.192&0.004	&un.sat&	11.654&0.027&	11.022&0.030	&10.847&0.018\\				
KIC 6183511	&283.4019&	41.5163	&17.194&0.006&	un.sat	&12.803&0.021&	12.231&0.024&	12.010&0.028\\
KIC 3426367	&285.9289&	38.5210	&15.960&0.008&	un.sat&	11.804&0.022&	11.193&	0.021&10.993&0.18\\
KIC 3935942&	285.2238&	39.0302&	17.165&0.009&	un.sat&	12.951&0.024	&12.294&0.027&	12.114&0.024\\										
KIC 4543236	&285.2670&	39.6315&	17.317&	0.013&un.sat&	13.279&0.027	&12.725	&0.033&12.478&0.030\\								
KIC 4243354	&285.4000	&39.3127	&16.412&0.007&w.sat&	13.092&0.022	&12.461&0.019&	12.272&0.023\\									
KIC 5513769&	284.2961	&40.7159	&16.708&0.008&	un.sat&	13.003&0.022&12.361&0.022&	12.148&0.020\\
KIC 4725681&	283.8664&	39.8981	&16.703&0.005&	un.sat	&13.182&0.024&	12.510&0.023&	12.295&0.022\\
KIC 4139816&	286.0791	&39.2783&	16.896&0.012	&w.sat&	13.863&0.026&	13.221&0.023&	13.080&0.030\\
J20515725\tablenotemark{\it{e}}	&312.9885&	$-1.1923$	&19.376&0.079&	un.sat&	15.712&0.087&	15.003&0.094&14.957& 0.145\\
KIC 4543619	&285.4578&	39.6321	&15.873&0.014&	un.sat	&13.153&0.023&	12.535&0.022&	12.411& 0.022\\				
KIC  2850521 &	291.0475	&38.0901	& 15.831 &0.014&	w.sat &	13.207&0.024&	12.581&0.022	& 12.456&0.024\\

  \enddata 
  \tablenotetext{\it{a}}{\small{The astrometry is taken from 2MASS observations.}}
  \tablenotetext{\it{b}}{The statistical photometric uncertainty in the {\it {g}}-band photometry, the systematic errors ($\sim 2\%$, Ivezi\'{c} et al.~2004 ) are not included.}
  \tablenotetext{\it{c}}{The saturation state of {\it{g}} band, un.sat = unsaturated and  weak.sat = weakly saturated}
 \tablenotetext{\it{d}}{The total photometric uncertainty, including the corrected band photometric uncertainty, the nightly photometric zero-point uncertainty and the flat-fielding residual error}
 \tablenotetext{\it{e}}{\small{2MASSJ20515725-0111317}}
 \tablecomments{\small {Name abbreviation: LJ=LSPM J}}f
  \tablecomments{\small {In all uncertainties above, the errors due to extinction corrections are not considered.}}
 
 \end{deluxetable}

\begin{deluxetable}{rrrrr}
\tabletypesize{\scriptsize}
 \placetable{t}
 \tablecolumns{5}
 \tablewidth{0pc}
 \tablecaption{Spectral Type and Metallicity  of 49 dwarf stars in the calibration sample}
 \tablehead{
 \colhead {Name} & \colhead {Spectral Type} & \colhead {Spectral Type Ref.} &  \colhead {[Fe/H]} & \colhead {[Fe/H]{Ref.}} }
 \startdata 
LSPM J0856+1239&	M6&	N14&	+0.07 $\pm$ 0.12&	N14\\
LSPM J1005+1703& 	M6&	N14&	+0.06 $\pm$ 0.12&	N14\\
LSPM J1530+0926& 	M6&	N14&	+0.09 $\pm$ 0.12&	N14\\
LSPM J2012+0112&	M6&	N14&	+0.06 $\pm$ 0.12&	N14\\
LSPM J0738+1829& 	M6&	N14&	+0.30 $\pm$ 0.12&	N14\\
LSPM J1002+4827&	M6&	N14&	+0.28 $\pm$ 0.12&	N14\\
LSPM J1432+0811&	M6&	N14&	+0.24 $\pm$	0.12&	N14\\
LSPM J1631+4051&	M6&	N14&	+0.30	$\pm$ 0.12&	N14\\
LSPM J0738+4925& 	M6&	N14&	+0.17	$\pm$ 0.13&	N14\\
LSPM J1239+0410&	M6&	N14&	+0.30	$\pm$ 0.12&	N14\\
LSPM J1309+2859&	M5&	N14&	+0.08	$\pm$ 0.13&	N14\\
LSPM J1021+0804& 	M5&	N14&	+0.10	$\pm$ 0.12&	N14\\
LSPM J1352+6649&	M5&	N14&	+0.15	$\pm$ 0.12&	N14\\
LSPM J0918+6037W&	M5&	N14&	+0.10	$\pm$ 0.12&	N14\\
LSPM J0001+0659& 	M5&	N14&	+0.10	$\pm$ 0.12&	N14\\
LSPM J0237+0021&	M5&	N14&	+0.03	$\pm$ 0.12&	N14\\
LSPM J0917+5825&	M5&	N14&	+0.08	$\pm$ 0.12&	N14\\
LSPM J1031+5705& 	M5&	N14&	+0.01	$\pm$ 0.13&	N14\\
LSPM J1112+0338&	M5&	N14&	+0.04	$\pm$ 0.13&	N14\\
LSPM J1419+0254&	M5&	N14&	+0.04	$\pm$ 0.13&	N14\\
LSPM J0024+2626&  	M5&	N14&	+0.29	$\pm$ 0.12&	N14\\
LSPM J1348+0429& 	M5&	N14&	+0.27	$\pm$ 0.12&	N14\\
LSPM J1101+0300&	M5&	N14&	$-0.30$	$\pm$ 0.13&	N14\\
NLTT 8870& M5&M14& +0.11 $\pm$ 0.10&  M14-13a Empir.Calib\\
LSPM J0959+4712&	M4&	N14&	+0.01 $\pm$	0.12&	N14\\
LSPM J1148+5305&	M4&	N14&	$-0.01$	$\pm$ 0.13&	N14\\
LSPM J1110+4757&	M4&	N14&	$-0.09$ $\pm$ 0.13&	N14\\
LSPM J0958+0558& 	M4&	N14&	+0.03 $\pm$	0.12&	N14\\
LSPM J1316+2752&	M4&	N14&	$-0.20$	$\pm$ 0.14&	N14\\
LSPM J1709+3909& 	M4&	N14&	$-0.23$ $\pm$ 0.14&	N14\\
LSPM J1240+1946& M4& N14& $-0.14$ $\pm$ 0.14& N14\\
LSPM J0920+0322&M4& N14& $-0.28$ $\pm$ 0.14& N14\\
LSPM J1345+2852&	M3.5&	D13/R10$-13$&	$-0.13$	$\pm$ 0.12&	D13\\
LP 655$-$23&	M3&	M14&	+0.03	$\pm$ 0.10&	M14\\
LSPM J1314+4846&	M3&	N14&	$-0.13$	$\pm$ 0.12&	N14\\
LSPM J1038+4831&  	M3&	N14&	+0.24	$\pm$ 0.12&	N14\\
LSPM J2148+0126 & M3& N14& +028 $\pm$ 0.12& N14\\
KIC 5252367&	M2&	P12/R10$-$13&	$-0.18$	$\pm$ 0.13&	M13b/Non-KOI sample\\
KIC 6183511&	M1.5&	P12/R10$-$13&	$-0.15$	$\pm$ 0.06&	M13b/KOI sample\\
KIC 3426367&	M1.5&	P12/R10$-$13&	$-0.15$	$\pm$ 0.13&	M13b/Non-KOI sample\\
KIC 3935942&	M1&	P12/R10$-$13&	+0.04	$\pm$ 0.05&	M13b/Non-KOI sample\\
KIC 4543236&	M1&	P12/R10$-$13&	$-0.19$	$\pm$ 0.06&	M13b/Non-KOI sample\\
KIC 4243354&	M0&	P12/R10$-$13&	$-0.33$	$\pm$ 0.12&	M13b/Non-KOI sample\\
KIC 5513769&	M0&	P12/R10$-$13	&$-0.05$ $\pm$ 0.09	&M13b/Non-KOI sample\\
KIC 4725681&	M0&	P12/R10$-$13&	$-0.01$	$\pm$ 0.07&	M13b/KOI sample\\
KIC 4139816&	K7.5&	P12&	$-0.51$	$\pm$ 0.07&	M13b/KOI sample\\
2MASSJ20515725\tablenotemark{\it{a}}&K7&	SSPP/ST	&$-0.58$ $\pm$ 0.05&	SSPP/M\\
KIC 4543619&	K6.5&	MQ14&	$-0.55$ $\pm$ 0.12&	M13b/Non-KOI sample\\
KIC  2850521&	K6	&	P12&$-0.68$	$\pm$ 0.11	&M13b/Non-KOI sample\\

\enddata 

\tablenotetext{\it{a}}{\small{2MASSJ20515725-0111317}}
\tablecomments{\small {\textbf{Source of Spectral types:} M14=Mann et al.~(2014), N14=Newton et al.~(2014), D13/R10$-$13=Temperature taken from Deshpande et al.~(2013) and spectral type determined by the temperature-spectral relations of R10 and R13, P12/R10$-$13=Temperatures taken from Pinsonneault et al.~(2012) and spectral type determined by the temperature-spectral relation of R10 and R13, P12= Spectral type estimated  by temperatures taken from Effective temperature scale for KIC stars (Pinsonneault et al.~2012), SSPP/ST= Spectral type estimated by temperature taken from the Sloan Extension for Galactic Understanding and Exploration (SEGUE) Stellar Parameter Pipeline (SSPP, Lee et al.~2008), and MQ14=Spectral type estimated by temperature taken from  Rotation periods of Kepler main sequence stars (McQuillan 2014)
\textbf{Source of [Fe/H]:} 
N14=Based on an empirical metallicity calibration of moderate-resolution, NIR spectra in Newton et al.~(2014), M13b/KOI sample=The sample taken from Table 1 in Mann et al.~(2013b) with metallicities based on a modified empirical calibration of moderate-resolution, optical spectra in Mann et al.~(2013a;2013b), M13b/Non-KOI sample= The sample taken from Table 2 in Mann et al.~(2013b) with metallicities based on the weighted means of {\it J}-, {\it H}-, and {\it K}-band calibrations described in Mann et al.~(2013a), and SSPP/M= Based on multiple approaches of  SSPP; M14-13a Empir.Calib=Mann et al.~(2014), based on the empirically spectroscopic calibration through moderate-resolution observed spectra;
Rob07=Robinson et al.~(2007) through moderate-resolution observed spectra}
}
 
\end{deluxetable}

\section*{REFERENCES}

\noindent
Adelman-McCarthy, J., et al.: 2006, ApJS 162, 38 

 \noindent
 Ahn, C. P., et al.~2012, ApJS, 203, 21 

 \noindent
Bessell, M. S. \& Brett, J. M. 1988, PASP, 100, 1134

  \noindent
 Bochanski, J. J., et al.~2010, AJ, 139, 2679

  \noindent
 Bochanski, J. J., et al.~2013, AJ, 145, 40
 
  \noindent
 Bonfils, X., et al.~2005, A\&A, 442, 635

\noindent
Carlberg, R. G., Dawson, P. C., et al, 1985, APJ, 294, 674  
  
 \noindent
 Casagrande, L., et al.~2011, A\&A, 530, A138

\noindent
Casuso, E. \& Beckman, J. E. 2004, A$\&$A, 419, 181
  
\noindent
Clayton, D. D.,~1985, APJ, 285, 411

\noindent
Clayton, D. D.,~1987, APJ, 315, 451

\noindent
Clayton, D. D.,~1988, MNRAS, 234, 1 

\noindent
Covey, K. R., et al.~2007, AJ, 134, 2398

\noindent
Covey, K. R., et al. 2008, AJ, 136, 1778

\noindent
Cutri, R. M., et al.~2003, The IRSA 2MASS All-Sky Point Source Catalog, NASA/IPAC Infrared Science Archive

\noindent
Deshpande, R., et al.~2013, AJ, 146, 156

 \noindent
 Dhital, S., et al.~2012, AJ, 143, 67

 \noindent
Fuhrmann, K.~2008, MNRAS, 384, 173

 \noindent
 Fukugita, M., et al.~1996, AJ, 111, 1748

  \noindent
 Gilmore, G. \& Reid, N. 1983, MNRAS, 202, 1025

  \noindent
 Gunn, J. E., et al.~1998, AJ, 116, 3040

 \noindent 
Ivezi\'{c}, Z., et al.~2004, AN 325, 583
 
 \noindent
 Ivezi\'{c}, Z., et al.~2007, AJ, 134, 973

 \noindent
 Johnson, J. A. \& Apps, K. 2009, ApJ, 699, 933

 \noindent
 Johnson, J. A., et al.~2012, AJ, 143, 111

\noindent
Jones, D. O., et al, 2011, AJ, 142, 44

 \noindent
 Juri${\acute{\mathrm{c}}}$, M., et al.~2008, AJ, 673, 864 
 
\noindent
K\"{o}ppen, J. \& Arimoto, N. ~1990, A$\&$A, 240, 22

 \noindent
 Larson, K. A. \& Whittet, D. C. B. 2005, AJ, 623, 897

\noindent
Lee, Y. S., et al.~2008, AJ, 136, 2022

\noindent
 L\'{e}pine, S., et al.~2003, ApJ, 585, L69

 \noindent
 L\'{e}pine, S., et al.~2007, ApJ, 669, 1235

 \noindent
 L\'{e}pine, S., et al.~2013, AJ, 145,102

 \noindent
 Mann, A. W., et al.~2013a, AJ, 145, 52

\noindent
Mann, A. W., et al. 2013b, AJ, 770, 43

\noindent
Mann, A. W., et al. 2014, AJ, 147, 160

\noindent
McQuillan, A., et al.~2014, ApJS, 211, 24

 \noindent
 Mo, H., et al.~2010, Galaxy Formation and Evolution (Cambridge University Press)
 
 \noindent
 Nagashima, M. \& Okamoto, T. 2006, AJ, 643, 863

 \noindent
 Neves, V., et al.~2012, A\&A, 538, A25

\noindent
Newton, E. R., et al. 2014, AJ, 147, 20

  \noindent
 Nordstr\"om, B. 2008, The Ages of Stars, Proceedings IAU Symposium, No.258

\noindent 
 Ostriker, J. B. \& Thuan, T. X. 1975, Ap. J. Lett., 201, L51
 
 \noindent
 Padmanabhan, N., et al. 2008, AJ, 674, 1217
 
 \noindent
 Pagel, B. E. J. \& Patchett, B. E. 1975, MNRAS, 172, 13 
 
 \noindent
 Pagel, B. E. J. 2009, Nucleosynthesis and Chemical Evolution of Galaxies, Second edition (Cambridge University Press)

 \noindent
Pinsonneault, M. H., et al. 2012, ApJS, 199, 30
 
 \noindent
Prantzos, N.~2007, eprint arXiv: 1101.2108v1

\noindent
Rajpurohit, A. S., et al.~2010, SF2A-2010: Proceedings of the Annual meeting of the French Society of Astronomy and Astrophysics

 \noindent
 Rajpurohit, A. S., et al.~2013, A\&A, 556, 14

\noindent
Ram\'{i}rez, I., et al. 2007, A\&A, 465, 271

\noindent
 Reid, I. N. \& Gizis, J. E. 1997, AJ, 114, 1992

\noindent
Robinson, S. E.,~et al. 2007, ApJS, 169, 430

\noindent
Rojas-Ayala, B., et al. 2010, ApJL, 720, L113
 
 \noindent
 Rojas-Ayala, B., et al.~2012, ApJ, 748, 93

\noindent
Santos, N. C.,  et al.~2002, A$\&$A, 392, 215

\noindent
Santos, N. C., et al. 2004, A\&A, 415, 1153

 \noindent
 Schlaufman, K. C. \& Laughlin, G. 2010, A$\&$A, 519, A105

\noindent
Schlegel, D. J., et al.~1998, ApJ, 500, 525

 \noindent
 Schmidt, M. 1963, ApJ, 137, 758

  \noindent
 Skrutskie, M. F., et al.~2006, AJ, 131, 1163
 
 \noindent
 Sommer-Larsen, J.~1991, MNRAS, 243, 468

\noindent 
 Terrien, R. C., et al. 2012, ApJL, 747, L38
 
 \noindent
 Tucker, D. L., et al. 2006, Astron. Nachr., 327, 821
 
 \noindent
 Twarog, B. A., 1980a, ApJS, 44, 1
 
 \noindent
 Twarog, B. A., 1980b, ApJ, 242, 242
 
 \noindent
 Valenti, J. A., \& Piskunov, N. E. 1996, A$\&$AS, 118, 595

 \noindent
 Valenti, J. A. \& Fischer, D. A. 2005, ApJS, 159, 141

  \noindent
 West, A. A., et al. 2006, AJ, 132, 2507

 \noindent
 West, A. A., et al.~2008, AJ, 135, 785

 \noindent
 West, A. A., et al.~2011, AJ, 141, 97
 
 \noindent
 West, A. A., et al~2014, AAS Meeting No.224, 404.04

 \noindent
 Woolf, V. M. \& Wallerstein, G. 2005, MNRAS, 356, 963

  \noindent
 Woolf V. M., et al.~2009, PASP, 121, 117

 \noindent
 Woolf, V. M. \& West, A. A. 2012, MNRAS, 422, 1489

 \noindent
 Wyse, R. F. G., \& Gilmore, G.~1995, AJ, 110, 2771

 \noindent
 York, D. G., et al.~2000, AJ, 120, 1579

 \noindent
 Zagury, F. 2006, MNRAS, 370, 1763

\vspace{5cm}

\end{document}